\documentclass[nofootinbib,floats,aps,superscriptaddress,preprint]{revtex4}
\usepackage{graphicx}
\usepackage{axodraw}

\newlength{\captsize}           \let\captsize=\small
\newlength{\captwidth}          
\newlength{\beforetableskip}    
\newcommand{\capt}[1]{\begin{minipage}{\captwidth}
              \let\normalsize=\captsize
              \caption[0]{#1}
              \end{minipage}\\ \vspace{\beforetableskip}}

\def\iso{\mathchoice{\cong}{\cong}{\isoS}{\cong}}
\def\isoS{\vbox{\baselineskip 0pt  \lineskip 0.5pt
    \ialign{$ \mathsurround=0pt  \scriptstyle \hfil ## \hfil $\crcr
        \sim \crcr = \crcr}}}

\newcommand\longdash{\hbox{\rm{\phantom{a}---\phantom{a}}}}
\newcommand{\mathbold}[1]{\mbox{\boldmath $\bf #1$}}

\def\Tr{{\rm Tr}}
\def\abar{{\bar a}}
\def\bbar{{\bar b}}
\def\cbar{{\bar c}}
\def\dbar{{\bar d}}
\def\ebar{{\bar e}}
\def\fbar{{\bar f}}
\def\gbar{{\bar g}}
\def\hb{{\bar h}}
\def\Lama{\Lambda_1}
\def\Lamb{\Lambda_2}
\def\Lamc{\Lambda_3}
\def\Lamd{\Lambda_4}
\def\Lame{\Lambda_5}
\def\Lamf{\Lambda_6}
\def\Lamg{\Lambda_7}
\def\Lamcde{\Lambda_{345}}
\def\lam{\lambda}
\def\lamT{\lambda_T}
\def\lamU{\lambda_U}
\def\lamV{\lambda_V}
\def\lamA{\lambda_A}
\def\lamF{\lambda_F}
\def\wtil{\widetilde}
\def\ben{\begin{enumerate}}
\def\een{\end{enumerate}}
\def\beq{\begin{equation}}
\def\eeq{\end{equation}}
\def\beqa{\begin{eqnarray}}
\def\eeqa{\end{eqnarray}}

\def\ifmath#1{\relax\ifmmode #1\else $#1$\fi}
\def\lsim{\mathrel{\raise.3ex\hbox{$<$\kern-.75em\lower1ex\hbox{$\sim$}}}}
\def\gsim{\mathrel{\raise.3ex\hbox{$>$\kern-.75em\lower1ex\hbox{$\sim$}}}}
\def\sect#1{section~\ref{#1}}
\def\sects#1#2{sections~\ref{#1} and \ref{#2}}

\def\eq#1{eq.~(\ref{#1})}
\def\Ref#1{ref.~\cite{#1}}
\def\Refs#1#2{refs.~\cite{#1} and \cite{#2}}

\def\eqs#1#2{eqs.~(\ref{#1}) and (\ref{#2})}
\def\eqpair#1#2{eqs.~(\ref{#1}), (\ref{#2})}
\def\eqst#1#2{eqs.~(\ref{#1})--(\ref{#2})}
\def\eqthree#1#2#3{eqs.~(\ref{#1}), (\ref{#2}) and (\ref{#3})}
\def\Eq#1{Eq.~(\ref{#1})}

\def\tto{\leftrightarrow}
\def\anti{\overline}

\def\irtwo{\nicefrac{1}{\sqrt{2}}}
\def\mud{M_U}
\def\mdd{M_D}
\def\vev#1{\langle #1 \rangle}
\def\qlo{Q^0_L}
\def\uro{U^0_R}
\def\dro{D^0_R}

\def\eiuo{\eta_1^{U,0}}
\def\eiiuo{\eta_2^{U,0}}
\def\eido{\eta_1^{D,0}}
\def\eiido{\eta_2^{D,0}}

\def\eiuoi{\eta_i^{U,0}}
\def\eidoi{\eta_i^{D,0}}
\def\eiui{\eta_i^U}
\def\eidi{\eta_i^D}
\def\kpuo{\kappa^{U,0}}
\def\rhuo{\rho^{U,0}}
\def\kpdo{\kappa^{D,0}}
\def\rhdo{\rho^{D,0}}
\def\kpu{\kappa^U}
\def\rhu{\rho^U}
\def\kpd{\kappa^D}
\def\rhd{\rho^D}
\def\Lam{\Lambda}

\def\cbma{c_{\beta-\alpha}}

\def\sbma{s_{\beta-\alpha}}
\def\ctwob{c_{2\beta}}
\def\stwob{s_{2\beta}}
\def\sthreeb{s_{3\beta}}

\def\cthreeb{c_{3\beta}}

\def\lamtil{\lam\ls{345}}
\def\lamhat{\widehat\lam}
\def\phm{\phantom{-}}

\def\beq{\begin{equation}}
\def\eeq{\end{equation}}
\def\ifmath#1{\relax\ifmmode #1\else $#1$\fi}
\def\calm{\mathcal{M}}

\def\call{\mathcal{L}}
\def\tb{t_{\beta}}

\def\sb  {s_{\beta}}
\def\cb  {c_{\beta}}
\def\stwob  {s_{2\beta}}
\def\ctwob  {c_{2\beta}}
\def\sa  {s_{\alpha}}
\def\ca  {c_{\alpha}}

\def\tanb{\tan\beta}
\def\cotb{\cot\beta}
\def\sinb{\sin\beta}
\def\cosb{\cos\beta}

\def\sinbma{\sin(\beta-\alpha)}
\def\cosbma{\cos(\beta-\alpha)}
\def\sbmaii{s^2_{\beta-\alpha}}
\def\cbmaii{c^2_{\beta-\alpha}}

\def\hl{h}
\def\ha{A}
\def\hh{H}
\def\hpm{{H^\pm}}

\def\go{G^0}

\def\lamtil{\lam_{345}}

\def\mha{m_{\ha}}
\def\mhl{m_{\hl}}
\def\mhh{m_{\hh}}
\def\mhpm{m_{\hpm}}

\def\mw{m_W}

\def\ls#1{\ifmath{_{\lower1.5pt\hbox{$\scriptstyle #1$}}}}
\def\lss#1{\ifmath{^{\,\lower2.5pt\hbox{$\scriptstyle #1$}}}}
\def\nicefrac#1#2{\hbox{${#1\over #2}$}}

\def\half{\ifmath{{\textstyle{1 \over 2}}}}

\def\quarter{\ifmath{{\textstyle{1 \over 4}}}}

\def\ie{{\it i.e.}}
\def\eg{{\it e.g.}}

\renewcommand{\Re}{{\rm Re}}
\renewcommand{\Im}{{\rm Im}}

\begin{document}
%
\preprint{
\vbox{\vspace*{2cm}
      \hbox{IPPP/03/23}
      \hbox{DCPT/03/46}
      \hbox{SCIPP-04/15}
      \hbox{hep-ph/0504050}
      \hbox{April, 2005}
}}
\vspace*{4cm}

\title{Basis-independent methods for the two-Higgs-doublet model}
\author{Sacha Davidson}
\affiliation{Institute for Particle Physics Phenomenology \\
   University of Durham, South Road, Durham DH1 3LE, U.K.
}
\affiliation{Institut de Physique Nucl\'eaire de Lyon \\
Universit\'e Claude Bernard Lyon 1 \\
4, rue Enrico Fermi, F-69622 Villeurbanne, France}

\author{Howard E. Haber$^{1,}$}
\affiliation{Santa Cruz Institute for Particle Physics  \\
   University of California, Santa Cruz, CA 95064, U.S.A. \\
\vspace{2cm}}

\begin{abstract}

In the most general two-Higgs-doublet model (2HDM), unitary
transformations between the two Higgs fields do not change the
functional form of the Lagrangian.  All physical observables of the
model must therefore be independent of such transformations (\ie,
independent of the Lagrangian basis choice for the Higgs fields). We
exhibit a set of basis-independent quantities that determine all
tree-level Higgs couplings and masses.  Some examples of the
basis-independent treatment of 2HDM discrete symmetries are
presented.  We also note that the ratio of the neutral Higgs field
vacuum expectation values, $\tanb$, is not a meaningful parameter in
general, as it is basis-dependent.  Implications for the more
specialized 2HDMs (\eg, the Higgs sector of the MSSM and the
so-called Type-I and Type-II 2HDMs) are explored.

\end{abstract}

\maketitle

\section{Introduction}  \label{sec:intro}

The Standard Model of electroweak physics, which posits a single
hypercharge-one, SU(2)$_{\rm L}$ doublet (complex) Higgs field, provides a
extremely successful description of observed electroweak phenomena.
Nevertheless, there are a number of motivations to extend the Higgs
sector of this model by adding a second complex doublet Higgs
field~\cite{Lee:1973iz,Peccei:1977hh,susyhiggs,hhg,Carena:2002es,branco}.
Perhaps the best motivated of these extended models is the minimal
supersymmetric extension of the Standard Model (MSSM)~\cite{susyreviews},
which requires
a second Higgs doublet (and its supersymmetric fermionic partners) in
order to preserve the cancellation of gauge anomalies.  The Higgs sector
of the MSSM is a two-Higgs-doublet model (2HDM), which contains two
Higgs supermultiplets $\widehat H_u$ and $\widehat H_d$,
that are distinguished by the
sign of their hypercharge~\cite{susyhiggs,hhg,Carena:2002es}.
This establishes an unambiguous
``theoretical'' basis for the Higgs sector of the Lagrangian.
The structure of the MSSM Higgs sector is constrained by the
supersymmetry, leading to numerous relations among Higgs masses and
couplings.  However, due to supersymmetry-breaking effects, all such
relations are modified by loop-corrections, where the effects of
supersymmetry-breaking can enter.  Thus, one can describe the
Higgs-sector of the (broken) MSSM by an effective field theory
consisting of the most general two-Higgs-doublet model.

In a realistic model, the Higgs-fermion
couplings must be chosen with some care~\cite{Weinberg,Georgi} to avoid
flavor-changing-neutral-currents (FCNC). The 2HDM
can be classified by how this issue is addressed.
In type-I models~\cite{type1,hallwise},
there exists a basis choice in which
only one of the Higgs fields couples to the Standard Model fermions.
In type-II models~\cite{type2,hallwise}, there exists a basis choice in which
one Higgs field couples to up-type quarks,
and the other Higgs field couples to down-type quarks.
Type-III models~\cite{typeiii} allow
both Higgs fields to couple to all the Standard
Model fermions; such a model is phenomenologically viable only if
the resulting FCNC-couplings are sufficiently small.

From a phenomenological bottom-up perspective, it is
important to study the properties of the most general
2HDM~\cite{Diaz:2002tp} without
imposing any special relations among the tree-level parameters.  The
2HDM Lagrangian depends on two
identical hypercharge-one, SU(2)$_{\rm L}$ doublet Higgs fields, $\Phi_1$ and
$\Phi_2$.  If $\Phi_1$ and $\Phi_2$ couple
identically to all other fields (fermions and gauge bosons), they
are only distinguished by the scalar interactions
contained in the scalar Higgs potential.
Consequently, one is free to choose an arbitrary
basis in the Lagrangian
for the two dimensional space of Higgs fields~\cite{silva},
corresponding to
two arbitrary orthogonal linear combinations of $\Phi_1$ and
$\Phi_2$.  The couplings and mass matrix elements will
depend on this basis choice.
In particular, the important parameter of the MSSM,
$\tan\beta\equiv \vev{H_u^0}/\vev{H_d^0}$, is not a physical parameter
of the most general 2HDM since its definition refers to a specific basis
choice for the Higgs fields.
The aim of this paper is to establish a basis-independent
formalism for analyzing the most general 2HDM.   This provides
a framework for
expressing any physical observable (\textit{e.g.}, Higgs masses and
couplings) in a form that is independent of the basis choice.

It is often phenomenologically desirable to express the parameters of
a field theory as functions of basis-independent
physical observables~\cite{jarlskog,cpinvariants,basisind}.
This procedure has various advantages: it leads to parameters that are
uniquely defined, gauge invariant and typically
well behaved under renormalization.
One simple and straightforward alternative is
to choose a basis and consistently calculate there. However,
results expressed in terms of basis-dependent Lagrangian
parameters are
meaningless without identifying the basis, and the
comparison with results computed
in a different basis choice is often difficult.
This situation can be avoided by
constructing quantities that are invariant
under arbitrary basis transformations in the space of fields,
and expressing the fundamental Lagrangian parameters in terms of these
``invariants''.  We wish to draw attention to this procedure
in the 2HDM, where one must specify
the basis for the scalar fields before defining the Higgs
potential parameters, the Higgs-fermion Yukawa couplings and
$\tan\beta$.

In the 2HDM, we define ``invariants'' to be quantities
that are scalar under arbitrary unitary
transformations among the two Higgs fields
in the Lagrangian~\cite{branco,silva,lavoura,Ginzburg:2004vp}.
All physical observables can be expressed in terms of invariants.
The parameters that appear in the Lagrangian with respect to a generic
basis of Higgs fields are not physical.  Nevertheless, even in the most
general model, certain specific basis choices are singled out, and the
corresponding Lagrangian parameters become meaningful. Henceforth, we
shall designate such parameters as ``physical parameters.'' For example,
in the CP-conserving 2HDM, the Lagrangian parameters in the
basis corresponding to the neutral CP-even Higgs mass eigenstates are
physical parameters~\cite{boudjema}. That is,
any coupling or mixing angle in the mass eigenstate basis can be
expressed in terms of invariants.
One particularly useful class of bases is the so-called
\textit{Higgs basis}~\cite{Donoghue:1978cj,Georgi,silva,lavoura,lavoura2}
in which only one of the two scalar fields exhibits a non-zero vacuum
expectation value.  We shall demonstrate that the Lagrangian parameters
with respect to the Higgs basis are closely related to physical parameters.

In this paper, we construct invariants in two ways.
In the body of the paper, we combine covariant tensors
constructed from the Higgs potential parameters and the vacuum
expectation values
to obtain scalar quantities. This elegant approach
is rather abstract, so in the appendices we explicitly
construct eigenbases corresponding to the orthonormal eigenvectors
of various physically relevant second-ranked tensors.
By contracting these eigenvectors with the Higgs potential parameter tensors,
one can identify which combinations are invariant and correspond to
physical parameters.

In \sect{sec:three}, the scalar potential
is expressed in a covariant notation with respect to the U(2)
transformations between the two Higgs fields,
and we introduce a set of invariants that govern
the model.  These invariants depend on the Higgs potential
parameter tensors and
a matrix $V$ constructed from
the vacuum expectations values.  The eigenvectors of $V$ form an
orthonormal basis that defines the Higgs basis.
A review of the 2HDM
in a generic basis and in the Higgs basis
can be found in Appendix \ref{sec:two}, where we also
exhibit explicit relations among the
corresponding Higgs potential parameters.
In Appendix \ref{app:three} we explicitly
combine eigenvectors of $V$ with Higgs potential parameter tensors
to obtain
quantities that depend linearly on the Higgs
potential parameters.
We show that  some of these are
invariant with respect to U(2) transformations
and hence ``physical'', whereas  others have
undetermined phases.
In \sect{sec:discrete},  we construct invariants
that  vanish when
the discrete symmetries $\Phi_1\to\Phi_1$, $\Phi_2 \rightarrow
- \Phi_2$ and CP-symmetry, respectively, are realized.
The relation of the
 $\Phi_1\to\Phi_1$, $\Phi_2 \rightarrow
- \Phi_2$ discrete symmetry  to the permutation symmetry
$\Phi_1\tto\Phi_2$ is discussed in Appendix \ref{app:B}.
In  \sects{sec:four}{sec:five}, we
restrict our analysis to the CP-conserving
2HDM scalar potential.  We first construct
the physical Higgs boson couplings
to vector bosons and the Higgs self-couplings
in terms of basis-independent
parameters.  We then study the Higgs-fermion
interactions in the 2HDM, using a choice of basis
motivated by the model to compute the physical Higgs-fermion
couplings in terms of invariant couplings.  In doing so,
new $\tan\beta$-like parameters arise that are directly defined
in terms of physical Yukawa couplings.  Both CP-conserving and
CP-violating Higgs-fermion interactions are treated.
Finally, a brief summary is provided in \sect{sec:six}.

\section{Basis-independent analysis of the 2HDM}
\label{sec:three}

The most general two-Higgs-doublet extension of the
Standard Model~\cite{hhg,Diaz:2002tp,ghdecoupling}
is a theory of two identical hypercharge-one complex doublet
scalar fields $\Phi_1$ and $\Phi_2$.  We shall denote the
complex two-dimensional space spanned by these two fields as the Higgs
flavor space.  The canonically normalized
(gauge-covariant) kinetic energy terms of the
scalar fields are invariant under arbitrary global U(2) transformations
in this flavor space.  Thus, we are free to redefine our two scalar
fields by making an arbitrary U(2) transformation.  A specific choice
of fields will be called a \textit{choice of basis}.  Since physical
observables are independent of this choice of basis, it is desirable
to formulate the Higgs sector of the theory in a basis-independent
manner.

Nearly all discussions of Higgs physics employ a specific basis
choice.  In the \textit{generic basis}, one expands the Higgs fields
around their vacuum expectation values $\vev{\Phi_a^0}=v_a/\sqrt{2}$
($a=1,2$), and the parameter $\tan\beta\equiv v_2/v_1$ plays an
important role in the Higgs phenomenology.  However, this parameter
cannot be meaningful in general, as it is basis-dependent.  In fact,
this parameter disappears completely if one transforms to the
\textit{Higgs basis}, where the Higgs vacuum expectation value resides
solely in one of the two Higgs doublets.\footnote{The generic basis and Higgs
basis are reviewed in Appendix \ref{sec:two}.  In particular,
\eqst{maa}{Lam7def} provide
the relations between the Higgs potential parameters in the
generic and Higgs bases.}
This suggests that the Higgs basis is special.  In this section, we shall
demonstrate that the parameters that appear in the Higgs basis are closely
related to  basis-independent quantities.

The scalar Higgs potential of the 2HDM can be
written in terms of the two complex hypercharge-one,
SU(2)$\ls{L}$ doublet scalar
fields,  following \Ref{branco}, as:
\beq \label{genericpot}
\mathcal{V}=Y_{a\bbar}\Phi_\abar^\dagger\Phi_b
+\half Z_{a\bbar c\dbar}(\Phi_\abar^\dagger\Phi_b)
(\Phi_\cbar^\dagger\Phi_d)\,,
\eeq
where the indices $a$, $\bbar$, $c$ and $\dbar$
run over the two-dimensional Higgs
flavor space and
\beq \label{zsym1}
Z_{a\bbar c\dbar}=Z_{c\dbar a\bbar}\,.
\eeq
Hermiticity of $\mathcal{V}$ implies that
\beqa
Y_{a \bbar}&=& (Y_{b \abar})^\ast\,, \label{ysym}\\
Z_{a\bbar c\dbar}&=& (Z_{b\abar d\cbar})^\ast\,. \label{zsym2}
\eeqa
We can match the
standard 2HDM notation given in \eq{pot} by
making the following identifications:
\beqa \label{ynum}
&& Y_{11}=m_{11}^2\,,\qquad\qquad \phantom{Y_{12}}\,\,\,
Y_{12}=-m_{12}^2 \,,\nonumber \\
&& Y_{21}=-(m_{12}^2)^\ast\,,\qquad\qquad Y_{22}=m_{22}^2\,,
\eeqa
and
\beqa \label{znum}
&& Z_{1111}=\lam_1\,,\qquad\qquad \,\,\phantom{Z_{2222}=}
Z_{2222}=\lam_2\,,\nonumber\\
&& Z_{1122}=Z_{2211}=\lam_3\,,\qquad\qquad
Z_{1221}=Z_{2112}=\lam_4\,,\nonumber \\
&& Z_{1212}=\lam_5\,,\qquad\qquad \,\,\phantom{Z_{2222}=}
Z_{2121}=\lam_5^\ast\,,\nonumber\\
&& Z_{1112}=Z_{1211}=\lam_6\,,\qquad\qquad
Z_{1121}=Z_{2111}=\lam_6^\ast\,,\nonumber \\
&& Z_{2212}=Z_{1222}=\lam_7\,,\qquad\qquad
Z_{2221}=Z_{2122}=\lam_7^\ast\,.
\eeqa

We assume that the vacuum respects the electromagnetic
gauge symmetry. That is, the vacuum expectation values
of $\Phi_1$ and $\Phi_2$ are assumed to
be aligned in $SU(2)_L$ space,
and we follow the standard convention (after using the appropriate
SU(2)$_{\rm L}$ transformation) in writing:
\beq \label{sachavev}
\langle \Phi_a
\rangle={v\over\sqrt{2}} \left(
\begin{array}{c} 0\\ \widehat v_a \end{array}\right)\,,
\eeq
where $\widehat v_a$ is a vector of unit norm.
Taking the derivative of \eq{genericpot} with respect to $\Phi_b$, and
setting $\vev{\Phi^0_a}=v_a/\sqrt{2}$, we find the covariant form for
the scalar potential minimum conditions:
\beq \label{potmingeneric}
\widehat v_\abar^\ast\,
[Y_{a\bar b}+\half v^2 Z_{a\bbar c\dbar}\, \widehat v_\cbar^\ast\,
\widehat v_d]=0 \,.
\eeq

Under the flavor SU(2),
the two Higgs doublets fields transform as
$\Phi_a\to U_{a\bbar}\Phi_b$
(and $\Phi_\abar^\dagger=\Phi_\bbar^\dagger U^\dagger_{b\abar}$),
where $U^\dagger_{b\abar}U_{a\cbar}=\delta_{b\cbar}$ and ${\rm det}~U=1$.
Likewise,
the tensors $Y$ and $Z$ transform covariantly: $Y_{a\bbar}\to U_{a\cbar}
Y_{c\dbar}U^\dagger_{d\bbar}$
and $Z_{a\bbar c\dbar}\to U_{a\ebar}U^\dagger_{f\bbar}U_{c\gbar}
U^\dagger_{h\dbar} Z_{e\fbar g\hb}$.  The use of barred indices is
convenient for keeping track of which indices transform with $U$ and
which transform with $U^\dagger$.  For example, in this notation,
$v_\abar^\ast=(v_a)^\ast$, which makes the starred superscript on
$v_\abar$ superfluous (although we will maintain it for
the sake of additional clarity).
These SU(2) transformations are
symmetries of the physics (\textit{not} of the Lagrangian),
in the sense that physical observables must not depend on
the arbitrary basis choice made in the Lagrangian.

The flavor-SU(2) group of transformations  can be enlarged to
U(2).\footnote{The flavor-U(2) transformations
were first applied to the 2HDM in \Ref{silva}.  A nice textbook
discussion is given in \Ref{branco}.  These transformations have
recently been exploited by \Ref{Ginzburg:2004vp} in a study of the
CP-violating 2HDM.}
In particular, the transformation laws noted above continue to
hold where $U$ is a U(2) matrix (and the condition ${\rm det}~U=1$
is removed).
It should be noted that
U(2)~$\iso$~SU(2)$\times$U(1)$_{\rm Y}$, where
the (global) hypercharge U(1)$_{\rm Y}$ transformation
is a symmetry of the Lagrangian, and thus has no effect on the
parameters of the Higgs potential, $Y_{a\bbar}$ and $Z_{a\bbar c\dbar}$.
It is useful to keep track of the U(1)$_{Y}$ transformation properties of
various quantities because physical observables must also be
U(1)$_{Y}$-invariant.  Thus, we shall focus on U(2)-invariant scalars,
keeping in mind the distinction that
U(1)$_Y$ is a symmetry of the Lagrangian, whereas the
flavor-SU(2) transformations change the
parameters of the Lagrangian,

We shall construct various U(2)-invariants below that depend on the
Higgs potential parameters $Y_{a\bbar}$ and $Z_{a\bbar c\dbar}$
and the vacuum expectation values $\widehat v_a$.
In particular, any U(2)-invariant that depends on
the vacuum expectation values must be a function of
$\widehat v_a\widehat v_\bbar^\ast$.  Thus we
follow \Ref{branco} in defining
the hermitian matrix\footnote{$V_{a\bbar}$
is not an independent tensor, since it
depends on $Y$ and $Z$ via the scalar potential minimum
conditions [\eq{potmingeneric}].  The latter can be rewritten as:
$(VY)_{a\bbar}+\half v^2 Z_{e\bbar c\dbar}V_{a\ebar}V_{d\cbar}=0$.
Nevertheless, there is no simple closed-form formula for $V_{a\bbar}$
in terms of $Y_{a\bbar}$ and $Z_{a\bbar c\dbar}$.  Consequently, we shall
construct invariants that depend explicitly on $V_{a\bbar}$ in what follows.}
\beq \label{vabdef}
V_{a\bbar}\equiv \widehat v_a \,\widehat v_\bbar^\ast\,.
\eeq
The matrix $V$ transforms covariantly with respect to U(2).  Note that
$V$ possesses two eigenvalues, 1 and 0, corresponding to
orthonormal eigenvectors $\widehat v_a$ and
\beq
\widehat w_a\equiv -\epsilon_{ab}\widehat v_\bbar^\ast\,,
\eeq
with the inverse relation $\widehat v^\ast_{\abar}=\epsilon_{\abar\bbar}
\,w_b$. Here, orthogonality with respect to the complex
two-dimensional Higgs flavor space means that
$\widehat v_\abar\, \widehat w_a^\ast=0$.  In defining
the unit vector $\widehat w_a$, we have
introduced the tensors $\epsilon_{ab}$ and ${\epsilon}_{\abar\bbar}$, with
$\epsilon_{12}=-\epsilon_{21}=1$ and $\epsilon_{11}=
\epsilon_{22}=0$.\footnote{Note that $\delta_{a\bbar}$ is a
flavor-U(2)-invariant tensor, whereas
$\epsilon_{ab}$ and $\epsilon_{\abar\bbar}$ are only
flavor-SU(2)-invariant tensors.
The identity $\epsilon_{ab}\epsilon_{\cbar\dbar}=
\delta_{a\cbar}\delta_{b\dbar}-\delta_{a\dbar}\delta_{b\cbar}$ relates
the two tensors.}
The normalized
eigenvectors are defined only up to an arbitrary phase,
which implies that invariant quantities can only depend on
$\widehat v$ and $\widehat w$ via the combination
$V_{a\bbar}$, as noted above.  In particular,
\beq \label{wabdef}
W_{a\bbar}\equiv \widehat w_a \widehat w_{\bbar}^\ast=\delta_{a\bbar}-
V_{a\bbar}
\eeq
so that $W_{a\bbar}$ is not an independent tensor.
\Eq{wabdef} and
the following results imply that $V$ and $W$ are projection operators:
\beq
V^2=V\,,\qquad W^2=W\,,\qquad VW=WV=0\,.
\eeq

Since the
tensors $V_{a\bbar}$, $Y_{a\bbar}$ and $Z_{a\bbar c\dbar}$ exhibit tensorial
properties with respect to global U(2) rotations in the Higgs flavor
space, one can easily construct invariants with respect to U(2) by
forming U(2)-scalar quantities.  Combinations of covariant tensors can
be represented diagrammatically.  The tensors
$V_{a\bbar}$, $Y_{a\bbar}$ and $Z_{a\bbar c\dbar}$ can be depicted
by two and four-point ``vertices'', with incoming (outgoing) lines
corresponding to unbarred (barred) indices, as shown in Fig.~1.
\begin{figure}[ht!]
\unitlength.5mm
\SetScale{1.418}
\begin{boldmath}
\begin{center}
\begin{picture}(130,45)(80,-10)
\ArrowLine(0,0)(30,0)
\ArrowLine(30,0)(60,0)
\Text(30,0)[c]{$\otimes$}
\Text(-3,0)[r]{$a$}
\Text(63,1)[l]{$\bar{b}$}
\Text(30,-12)[c]{$V_{a \bar{b}}$}
\ArrowLine(120,0)(150,0)
\ArrowLine(150,0)(180,0)
\Text(150,0)[c]{$\times$}
\Text(117,0)[r]{$a$}
\Text(183,1)[l]{$\bar{b}$}
\Text(150,-12)[c]{$Y_{a \bar{b}}$}
\ArrowLine(230,25)(255,0)
\ArrowLine(255,0)(280,25)
\ArrowLine(280,-25)(255,0)
\ArrowLine(255,0)(230,-25)
\Text(255,0)[c]{$\bullet$}
\Text(223,23)[l]{$a$}
\Text(287,23)[r]{$\bar{b}$}
\Text(287,-23)[r]{$c$}
\Text(222,-21)[l]{$\bar{d}$}
\Text(275,-2)[c]{$Z_{a \bar{b} c \bar{d}}$}
\end{picture}
\end{center}
\end{boldmath}
\vspace{10mm}
\capt{Diagrammatic representation of covariant tensors.  The two point
vertices $V_{a\bbar}$ and
$Y_{a\bbar}$ are indicated by the symbols $\mathbold{\otimes}$ and
$\mathbold{\times}$, respectively.  The four-point vertex
$Z_{a\bbar c\dbar}$ is depicted by four incoming line segments
(meeting at the vertex point) where the indices appear in
\textit{clockwise} order.
Unbarred (barred) indices are represented
by incoming (outgoing) directed line segments.}
\label{fig:vertex}
\end{figure}

As noted above, $\delta_{a\bbar}$
is the only U(2)-invariant tensor that can be used
to contract a pair of indices.  Hence, one can only contract an
unbarred index with a barred index (or vice versa).  For example,
\beq \label{zabdef}
Z^{(1)}_{a\dbar}\equiv \delta_{b\cbar}Z_{a\bbar c\dbar}=Z_{a\bbar b\dbar}\,,
\qquad\qquad
Z^{(2)}_{c\dbar}\equiv \delta_{b\abar}Z_{a\bbar c\dbar}=Z_{a\abar c\dbar}\,.
\eeq
Using the vertex rules given in Fig.~1, the two tensors of \eq{zabdef}
can be depicted
diagrammatically as the one-loop bubble diagrams shown in Fig.~2.
\begin{figure}[ht!]
\unitlength.5mm
\SetScale{1.418}
\begin{boldmath}
\begin{center}
\begin{picture}(60,50)(80,-10)
\ArrowLine(0,0)(30,0)
\ArrowLine(30,0)(60,0)
\ArrowArcn(30,15)(15,270,-90)
\Text(-2,0)[r]{$a$}
\Text(62,0)[l]{$\bar{c}$}
\Text(30,0)[c]{$\bullet$}
\Text(150,20)[c]{$\bullet$}
\Text(30,-8)[c]{$Z_{a \bar{b} b \bar{c}}$}
\ArrowArc(150,5)(15,-270,90)
\ArrowLine(120,20)(150,20)
\ArrowLine(150,20)(180,20)
\Text(118,20)[r]{$b$}
\Text(182,20)[l]{$\bar{c}$}
\Text(150,28)[c]{$Z_{a \bar{a} b \bar{c}}$}
\Text(15,-30)[l]{\textbf{(a)}~$Z^{(1)}_{a\cbar}$}
\Text(135,-30)[l]{\textbf{(b)}~$Z^{(2)}_{b\cbar}$}
\end{picture}
\end{center}
\end{boldmath}
\vspace{10mm}
\capt{The one-loop bubble diagrams corresponding to
(a) $Z^{(1)}_{a \bar{c}} = Z_{a \bar{b} b \bar{c}}$
and (b) $Z^{(2)}_{b \bar{c}}=  Z_{a \bar{a} b \bar{c}}$.
These two diagrams are distinguished, since the $Z$-vertex must be read in
a clockwise fashion.}
\label{fig:bubble}
\end{figure}
Similarly, U(2)-invariant
quantities can be represented by ``vacuum'' diagrams with no
external lines.

We now turn to the task of constructing
independent invariant (scalar) combinations of
Higgs potential parameters.  Such invariant quantities can be directly
related to physical quantities which must be basis-independent.
We may combine the tensors $Y_{a\bbar}$, $Z_{a\bbar c\dbar}$,
and $V_{a \bbar}$ to create scalar
quantities by summing over pairs of indices (following the rule that
the summed index pairs must contain one unbarred and one barred index).
It is also convenient to employ $W_{a\bbar}\equiv
\delta_{a\bbar}-V_{a\bbar}$ introduced above.
We can exhibit thirteen invariants that
depend on the Higgs potential parameters and the vacuum
expectation value by defining
six real flavor-U(2)-invariants
(denoted below by $Y_{1,2}$ and $Z_{1,2,3,4}$)
and four complex index-free quantities,
denoted by $Y_3$ and $Z_{5,6,7}$,
which are constructed from $Y_{a\bbar}$, $Z_{a\bbar c\dbar}$ $v_a$
and $w_a$.  These quantities
are invariant under the Higgs flavor-SU(2) transformations, but are not
invariant with respect to U(1)$_{\rm Y}$
as shown in Appendix~\ref{app:three}.
From these four complex SU(2)-``invariants'',
we may extract seven flavor-U(2)
invariants consisting of the magnitudes and relative phases of these
four complex quantities.

The simplest invariants that can be constructed with
$Y$, $V$ and $W$ are given by
\beqa
Y_1 &\equiv & \Tr~(YV)\,,\label{syvv1}\\
Y_2&\equiv& \Tr~(YW)\,,\label{syvv2}\\
|Y_3|^2 &\equiv &  \Tr~(VYWY)\,.\label{syvv3}
\eeqa
Likewise, the simplest invariants that
can be constructed with $Z$, $V$ and $W$ are given by:
\beqa
Z_1&\equiv& Z_{a\bbar c\dbar}\,V_{b\abar}V_{d\cbar}\,,\label{szvv1}\\
Z_2&\equiv& Z_{a\bbar c\dbar}\,W_{b\abar}W_{d\cbar}\,,\label{szvv2}\\
Z_3 &\equiv& Z_{a\bbar c\dbar}\,V_{b\abar}W_{d\cbar}\,,\label{szvv3}\\
Z_4 &\equiv & Z_{a\bbar c\dbar}\,W_{d\abar}V_{b\cbar}\,,\label{szvv4}\\
|Z_5|^2 &\equiv&
Z_{a\bbar c\dbar}Z_{e \fbar g \bar h}\,
V_{f \abar}  W_{b \ebar} V_{h \cbar}  W_{d \gbar} \,,\label{szvv5}\\
|Z_6|^2 &\equiv&Z_{a\bbar c\dbar}Z_{e \fbar g \bar h}\,
V_{f \abar}  V_{b \ebar} V_{h \cbar}  W_{d \gbar}\,,\label{szvv6}\\
|Z_7|^2 &\equiv& Z_{a\bbar c\dbar}Z_{e \fbar g \bar h}\,
V_{f \abar}  W_{b \ebar} W_{h \cbar}  W_{d \gbar} \,.\label{szvv7}
\eeqa
The remaining three invariants correspond to three relative phases of
$Y_3$ and $Z_{5,6,7}$.  One possible choice is:
\beqa
Y_3^2 Z_5^\ast &\equiv & (VYW)_{b\abar}(VYW)_{d\cbar}Z_{a \bbar c \dbar}
\,,\label{syz35} \\
Y_3 Z_6^\ast &=& (VYW)_{b\cbar} V_{d\abar}Z_{a \bbar c \dbar}
\,,\label{syz36} \\
Y_3 Z_7^\ast &=& (VYW)_{b\cbar} W_{d\abar}Z_{a \bbar c \dbar}
\,.\label{syz37}
\eeqa

Using the rules of Fig.~1,
each of the invariants above is easily represented by a
simple vacuum diagram (or sum of such diagrams).  Using these
diagrammatic techniques, one can easily show that no additional
independent invariants exist that are linear or quadratic in the Higgs
potential parameters.
Further higher-order invariants could be constructed containing additional
powers of $Y$, $V$ or $W$. However, we will
argue below, and prove in Appendix \ref{app:three} that any higher order
invariant can be expressed in terms of those given in
\eqst{syvv1}{syz37}.

The invariants given in \eqst{syvv1}{syz37} are particularly simple in
the Higgs basis [see Appendix \ref{sec:two}], where $V$ and $W$ are given by:
\beq \label{vwhiggs}
V=\left(\begin{array}{cc} 1&\quad 0 \\ 0 &\quad 0\end{array}\right)\,,
\qquad\qquad
W=\left(\begin{array}{cc} 0&\quad 0 \\ 0 &\quad 1\end{array}\right)\,.
\eeq
Following Appendix~\ref{sec:two}, we denote the Higgs potential parameters
in the Higgs basis by $M_{ij}^2$ $(i,j=1,2)$ and $\Lambda_k$ $(k=1,\ldots,7)$.
After inserting the Higgs basis forms for $V$ and $W$
into \eqst{syvv1}{syz37}, we obtain:
\beq
Y_1=M_{11}^2\,,\qquad Y_2=M_{22}^2\,,\qquad Z_i=\Lambda_i~(i=1,2,3,4)\,,
\eeq
for the real Higgs basis parameters and
\beqa
&&|Y_3|^2=|M_{12}^2|\,,\qquad |Z_i|^2=|\Lambda_i|^2\quad (i=5,6,7)\,,
\nonumber \\
&& Y_3^2 Z_5^\ast=[M_{12}^2]^2 \Lambda_5^\ast\,,\qquad
Y_3 Z_i^*=-M_{12}^2\Lambda_i^\ast\quad (i=6,7)\,,
\eeqa
for the complex Higgs basis parameters.

The form of \eq{vwhiggs} implies that $V$ and $W$ serve as projection
operators for the Higgs basis field invariants $H_1^\dagger H_1$ and
$H_2^\dagger H_2$. In particular, in the diagrammatic representation
of invariants, the insertion of $V$ projects out $H_1$ and the
insertion of $W$ projects out $H_2$.  Thus, each of the lines of the vacuum
diagrams corresponding to the invariants listed in \eqst{syvv1}{syz37}
can be identified with one of the two fields $H_1$ and $H_2$ of the Higgs
basis.  All such vacuum diagrams must contain an equal number of
fields and their complex conjugates (corresponding to ingoing and
outgoing lines).  Hence, there are no complex
U(2)-invariants that are linear in the scalar potential
parameters.

The scalar minimum conditions can be rewritten
in terms of the U(2)-invariants introduced above.  First we multiply
\eq{potmingeneric} by $\widehat v_b$ to obtain one of the minimum conditions:
\beq \label{potmininv1}
Y_1=-\half Z_1 v^2\,.
\eeq
A second minimum condition can be obtained by multiplying
\eq{potmingeneric} by $\widehat w_b$.  Using \eqs{yvv}{zvv6}, it follows
that $Y_3=-\half Z_6 v^2$, which can be rewritten in terms of
U(2)-invariant quantities as
\beq \label{potmininv2}
|Y_3|^2=\quarter |Z_6|^2 v^4\,,\qquad\qquad Y_3 Z_6^*=-\half |Z_6|^2 v^2\,.
\eeq
Consequently, after imposing the scalar minimum conditions, two of the
thirteen invariants above can be eliminated
via \eq{potmininv2}, leaving eleven
independent invariants.\footnote{Since we count $v^2$ as one of the
parameter degrees of freedom, the application of the scalar
potential minimum conditions does not affect the
overall counting of degrees of freedom.}
This corresponds precisely to the number of
parameter degrees of freedom of the 2HDM, as discussed in
Appendix \ref{sec:two}.

Having enumerated the independent invariants, one can begin to
relate them to physical observables of the theory.
One consequence of the analysis relating invariants to the Higgs
potential parameters in the Higgs basis is that the parameter
$\tanb$ is redundant (as it does not appear in
the Higgs basis description of the 2HDM).   More precisely, in the general
2HDM, $\tanb$ is a basis-dependent quantity.
This means that  physical quantities (\eg,
Higgs masses and physical couplings)
cannot depend on the (unphysical) angle between the
two vacuum expectation values.  Physical quantities can depend on
the angle between the vacuum expectation value $\widehat v_a$ and
another direction in the Higgs flavor space
picked out by an interaction ($e.g.$ a mass matrix
eigenvector).  We
illustrate this explicitly in \sect{sec:four}.
This suggests a procedure to provide a meaningful
definition of $\tan \beta$.
For example, even in the most general 2HDM, the Higgs
potential depends on tensorial quantities, which can be used to define
various matrices.  The two eigenvectors of these matrices can be used
as basis vectors which defines a basis that generally differs from the
Higgs basis, and thus can be used to define $\tanb$.
In a CP-invariant theory, an obvious choice would be the CP-even
mass eigenstate basis.

Other possible matrices, whose eigenvectors
can be used to define a new basis, include:
$Y_{a\bbar}$, $Z^{(1)}_{a\bbar}$ and $Z^{(2)}_{a\bbar}$.
However, in such cases there is a two-fold ambiguity in the
definition of $\tanb$ corresponding to the interchange of the two
identical scalar fields.  As an example, consider a definition of
$\tanb$ corresponding to a basis in which $Y_{12}=0$.  This is the
basis spanned by the eigenvectors of $Y$.  In this basis, we define
$|\widehat v_1|=\cos\beta$ and $|\widehat v_2|=\sin\beta$ (where
$0\leq\beta\leq\pi/2$).  A simple computation in the $Y_{12}=0$ basis yields:
\beqa
2~\Tr~(VY)-\Tr~Y&=&(Y_{11}-Y_{22})\cos 2\beta \,,\\[5pt]
2~\Tr~(Y^2)-(\Tr~Y)^2 &=&(Y_{11}-Y_{22})^2\,.
\eeqa
Thus, we arrive at one possible invariant definition of $\cos^2 2\beta$:
\beq \label{invbeta}
\cos^2
2\beta=\frac{\left[2~\Tr~(VY)-\Tr~Y\right]^2}{2~\Tr~(Y^2)-(\Tr~Y)^2}\,.
\eeq
The sign of $\cos 2\beta$ is ambiguous, and the ambiguity corresponds
to $\beta\to\pi/2-\beta$ (\textit{i.e.}, the interchange of the two
Higgs fields).

\section{Basis-Independent Description of Discrete Symmetries}
\label{sec:discrete}

In \sect{sec:three}, we assumed
that the scalar Higgs potential takes on the
most general possible form.  However, in some two-Higgs doublet
models, discrete symmetries are imposed that constrain the structure
of the Higgs potential.  Typically, the discrete symmetries are
formulated with respect to a particular basis, in which case the
symmetry is \textit{manifest}.   In a generic basis,
the discrete symmetry is of course still present, but in most cases it is
well disguised and not immediately evident.  Hence,
for a given discrete symmetry, it is desirable
to establish a basis-independent characterization.
In particular, it is especially useful to establish basis-independent
conditions that depend only on the $Y$ and $Z$ tensors, \textit{i.e.}
they do not require a determination of the vacuum expectation values
via the scalar potential minimum conditions.
A number of examples will be presented in sections
\ref{sec:discreteB} and \ref{sec:discreteC},
in which we identify invariant conditions for discrete symmetries.
These conditions can be evaluated in the Higgs basis, and provide a
connection to related results first obtained in \Ref{lavoura2}.

However, in some special regions of the Higgs potential parameter space
many of the invariant conditions that we obtain
are automatically satisfied (due
to the enhanced symmetry of the parameter region).
This complicates the search
for invariant conditions for the discrete symmetries,
so we first examine the nature of these exceptional regions of the
parameter space.

\subsection{An exceptional region of the Higgs potential parameter space}
\label{sec:discreteA}

Consider the explicit forms of $Z^{(1)}$ and $Z^{(2)}$
defined in \eq{zabdef}:
\beq \label{zonetwo}
Z^{(1)}=\left(\begin{array}{cc}\lam_1+\lam_4\quad &
\lam_6+\lam_7 \\ \lam_6^*+\lam_7^* \quad & \lam_2+\lam_4\end{array}\right)
\,,\qquad\qquad
Z^{(2)}=\left(\begin{array}{cc}\lam_1+\lam_3 \quad& \lam_6+\lam_7 \\
\lam_6^*+\lam_7^* \quad & \lam_2+\lam_3\end{array}\right)\,.
\eeq
Note that $Z^{(1)}$ and $Z^{(2)}$ are hermitian matrices that
commute so that they can be simultaneously diagonalized by
a unitary matrix.  It therefore follows that there
exists a basis in which $Z^{(1)}$ and $Z^{(2)}$ are simultaneously
diagonal; that is, $\lam_7=-\lam_6$.  This observation will be crucial
to many of the considerations of this section.

The existence of a basis in which $\lam_7=-\lam_6$ provides another
opportunity for checking the number of 2HDM independent parameters.
Once this basis is achieved, one can always rephase one of the Higgs
fields such that $\lam_6$ and $\lam_7$ are real.  This leaves
seven real parameters ($m_{11}^2$, $m_{22}^2$, $\lam_{1,2,3,4,6}$) and
two complex parameters ($m_{12}^2$ and $\lam_5$) for  a total of
eleven independent parameters.

The region of the Higgs potential parameter space where $\lam_1=\lam_2$ and
$\lam_7=-\lam_6$ is especially noteworthy.  In this case, $Z^{(1)}$
and $Z^{(2)}$ are both proportional to the $2\times 2$ unit matrix.
This must be true for \textit{all} basis choices.  Hence
if $\lam_1=\lam_2$ and $\lam_7=-\lam_6$ holds in one basis then it
holds in all bases.  A (basis-independent)
invariant condition can be given for this
case.  Consider
\beq
2~\Tr~[Z^{(1)}]^2-(\Tr~Z^{(1)})^2=(\lam_1-\lam_2)^2+4|\lam_6+\lam_7|^2\,.
\eeq
If this invariant quantity vanishes, then $\lam_1=\lam_2$ and
$\lam_7=-\lam_6$.

It is important to emphasize that the latter
represents a \textit{region} of Higgs parameter space.
That is, in this region of the parameter space,
any change of basis will generally modify the Higgs
potential parameters, subject to the condition that
$\lam_1=\lam_2$ and $\lam_7=-\lam_6$.  Two results will prove useful
in the following.  First, if $\lam_1=\lam_2$ and $\lam_7=-\lam_6$,
then one can always find a basis in which all the $\lambda_i$ are
real.\footnote{A simple phase redefinition can render $\lambda_6$ and
$\lambda_7$ real.  The nontrivial part of the proof, which is given in
\Ref{cpbasis}, demonstrates that a basis always exists in which
$\lam_5$, $\lam_6$ and $\lam_7$ are simultaneously real.}
Second, if $\lam_1=\lam_2$ and $\lam_7=-\lam_6$, then there exists a
basis in which $\lam_5$ is real and
$\lam_6=\lam_7=0$.  The proof of the latter result is
simple given the former result.  First, transform to a basis in which
all the $\lam_i$ are real.  Then, make one further U(2) transformation
given by \eq{utransform} with $\theta=\pi/4$ and $\gamma+\zeta=\pi/2$.  It is
easy to check that in the final basis, $\lam_6=\lam_7=0$.  Finally,
since $\lam_5$ is the only potentially complex parameter remaining, it
can be rendered real with a phase redefinition of one of the two Higgs
fields.

If $Y_{11}=Y_{22}$ and $Y_{12}=0$ then
$Y$ is proportional to the  $2\times 2$ unit matrix.
Again, one may conclude that if $Y_{11}=Y_{22}$ and $Y_{12}=0$ holds in one
basis then it holds in all bases.
Combining the two special cases just considered yields an
exceptional region of the Higgs potential parameter space in which
$Y_{11}=Y_{22}$, $Y_{12}=0$, $\lam_1=\lam_2$ and $\lam_7=-\lam_6$.
A possible discrete symmetry, which is respected by the scalar
Lagrangian and characterizes this exceptional choice of
parameters, is:
\beq \label{dissym}
\Phi_1\to e^{i\psi}\Phi_2^*\,,\qquad\qquad
\Phi_2\to -e^{i\psi}\Phi_1^*\,,
\eeq
where $\psi$ is an arbitrary phase.\footnote{Since
\eq{dissym} involves the transformation of
the scalar fields to their complex conjugates, it follow that this
discrete symmetry incorporates a charge conjugation transformation.}
It is straightforward to show that if \eq{dissym} is a symmetry in one
basis, then it is a symmetry in \textit{all} bases.  The results
above also imply that within the exceptional region of parameter
space, a basis exists in which  $Y_{a\bbar}\propto\delta_{a\bbar}$,
$\lam_1=\lam_2$, $\lam_5$ is real and $\lam_6=\lam_7=0$.

Alternatively, the exceptional region can be
characterized by two simultaneous ordinary $\mathbold{Z_2}$
symmetries.  In particular, consider a basis in which the scalar
Lagrangian respects the two discrete symmetries:
(i)~$\Phi_1 \rightarrow \Phi_1$, $ \Phi_2 \rightarrow - \Phi_2$
[to be discussed further in \sect{sec:discreteB}] and
(ii)~$ \Phi_1 \rightarrow \Phi_2$, $\Phi_2 \rightarrow - \Phi_1$.
Symmetry~(i) implies that $Y_{12} = 0$ and $\lambda_6 = \lambda_7 =
0$, while symmetry~(ii) implies that
$Y_{11} = Y_{22}$, $Y_{12}^*=-Y_{12}$, $\lam_5^*=\lam_5$,
and $\lambda_6 = - \lambda_7^*$.  If symmetries~(i) and (ii) are
simultaneously satisfied (in the same basis),
then $Y_{a\bbar}\propto\delta_{a\bbar}$, $\lam_1=\lam_2$, $\lam_5$ is real
and $\lam_6=\lam_7=0$, which indeed corresponds to the exceptional region of
parameter space in a particular basis.

\subsection{The discrete symmetry $\Phi_1\to\Phi_1$, $\Phi_2\to -\Phi_2$}
\label{sec:discreteB}

We consider first the discrete symmetry $\Phi_1\to
+\Phi_1$, $\Phi_2\to -\Phi_2$, which is realized
for some choice of basis.  This
discrete symmetry implies that $m_{12}^2=\lam_6=\lam_7=0$.
It is a discrete $\mathbold{Z}_2$ subgroup of the
Peccei-Quinn U(1) symmetry
$\Phi_1\to e^{i\alpha} \Phi_1$,
$\Phi_2\to e^{-i\alpha}\Phi_2$.\footnote{If $\lambda_5=0$ in the same
basis where the discrete $\mathbold{Z}_2$ symmetry is realized, then
the scalar Lagrangian respects the full Peccei-Quinn U(1) global
symmetry~\cite{Peccei:1977hh}.}
The basis-independent conditions for the $\Phi_1\to
+\Phi_1$, $\Phi_2\to -\Phi_2$ discrete symmetry
can be expressed in terms of commutators of matrices constructed from
the Higgs potential parameter tensors $Y$ and $Z$.
An alternate approach in terms of invariants constructed from
$Z$ and the eigenvectors of $Y$
is discussed in Appendix \ref{app:three},
along with the explicit eigenvector construction.

Our strategy for deducing the relevant commutator conditions is as
follows.  We first transform to a basis where
$\lam_7=-\lam_6$ [this is always
possible as discussed below \eq{zonetwo}].  In this basis, we search
for commutators that vanish when $Y_{12}=\lam_6=0$.  We then conclude
that the vanishing of such commutators implies that some basis must
exist where $m_{12}^2=\lam_6=\lam_7=0$.  We shall
make use of the
following two matrices:\footnote{Diagrammatically, $Z^{(11)}$
corresponds to a bubble-on-a-bubble with two legs (\textit{i.e.}, a
snowman) and $Y^{(1)}$ corresponds to a bubble on two legs with a cross
at the top of the bubble.}
\beq
Z^{(11)}_{c\dbar}\equiv Z^{(1)}_{b\abar}Z_{a\bbar c\dbar}\,,\qquad\qquad
Y^{(1)}_{c\dbar}\equiv Y_{b\abar}Z_{a\bbar c\dbar}\,.
\eeq
Then, consider the following two commutators evaluated in a basis
where $\lam_7=-\lam_6$:
\beqa
[Z^{(1)},Y]&=& (\lam_1-\lam_2)\left(\begin{array}{cc} 0& \quad Y_{12}
    \\ -Y_{12}^* & \quad 0\end{array}\right)\,,\label{comm1}\\[8pt]
[Z^{(1)},Z^{(11)}]&=& (\lam_1-\lam_2)^2\left(\begin{array}{cc} 0&
    \quad \lam_6 \\ -\lam_6^* & \quad 0\end{array}\right)
     \,,\label{comm2}
\eeqa

First we assume that $\lam_1\neq\lam_2$ in the $\lam_7=-\lam_6$ basis.
In this case, we find that
\beq \label{commcond}
[Z^{(1)},Y]=[Z^{(1)},Z^{(11)}]=0
\eeq
is the basis independent condition that guarantees the existence of a
basis in which $m_{12}^2=\lam_6=\lam_7=0$.
As an example, in
Appendix~\ref{app:B}, we demonstrate that if a 2HDM scalar potential
respects a permutation symmetry $\Phi_1\tto\Phi_2$ in some basis, then
\eq{commcond} is satisfied.  This result implies that
in some other basis the discrete symmetry
$\Phi_1\to\Phi_1$, $\Phi_2\to -\Phi_2$ must be manifest.

If $\lam_1=\lam_2$ and $\lam_7=-\lam_6$,
then $Z^{(1)}_{a\bbar}$, $Z^{(11)}_{a\bbar}\propto\delta_{a\bbar}$, and
\eqs{comm1}{comm2} automatically vanish.  In this case,
we are free to transform to a basis in which $Y$ is diagonal.
The following commutator is now relevant
(for $\lam_7=-\lam_6$ and $Y_{12}=0$):
\beq
[Y^{(1)},Y]= (Y_{11}-Y_{22})^2\left(\begin{array}{cc} 0& \quad -\lam_6
    \\ \lam_6^* & \quad 0\end{array}\right)\,.\label{comm3}
\eeq
Assuming that $Y_{11}\neq Y_{22}$, we find that:
\beq
[Y^{(1)},Y]=0\quad {\rm and} \quad Z^{(1)}_{a\bbar}\propto\delta_{a\bbar}
\eeq
are basis independent conditions that guarantee the existence of a
basis in which $m_{12}^2=\lam_6=\lam_7=0$.

None of the above arguments apply to the
exceptional region of parameter space [see \sect{sec:discreteA}] where
$Y_{11}=Y_{22}$, $Y_{12}=0$, $\lam_1=\lam_2$ and $\lam_7=-\lam_6$.
In this case, all three commutators [\eqthree{comm1}{comm2}{comm3}]
automatically vanish since all the matrices involved are proportional
to the unit matrix.  However, as noted at the end of \sect{sec:discreteA},
in the exceptional region, there exists a basis in which the
$\mathbold{Z_2}$ discrete symmetry $\Phi_1\to\Phi_1$, $\Phi_2\to -\Phi_2$
is manifest.   Thus, no further invariant conditions are required.

Finally, in the basis where the discrete symmetry is manifest
($m_{12}^2=\lam_6=\lam_7=0$),
\eq{cpx} implies that $\Im(\lam_5 e^{2i\xi})=0$.  One can
always choose $\lam_5$ real (and hence $\xi=0$) by redefining the phase of
$\Phi_2$.  We conclude that the Higgs scalar potential of this model
is CP-conserving.  Further considerations of CP invariance will be given in
\sect{sec:discreteC}.

We can generalize the results of this section in two ways.  First,
if there exists a basis in which $\lambda_6=\lambda_7=0$ but
$m_{12}^2\neq 0$, then the $\mathbold{Z}_2$ discrete symmetry is
\textit{softly} broken~\cite{Ginzburg:2004vp,lavoura2}.  A basis-independent
characterization of the softly-broken $\mathbold{Z}_2$ discrete symmetry
is:
\beq
[Z^{(1)},Z^{(11)}]=0\quad {\rm and}\quad [Z^{(1)},Y]\neq 0\,,
\eeq
assuming that $\lam_1\neq\lam_2$ in a basis
where $\lam_7=-\lam_6$.  In the case of
$\lam_1=\lam_2$ \textit{and} $\lam_7=-\lam_6$,
we proved in \sect{sec:discreteA}
that one can always transform to a new basis in which $\lam_5$ is real and
$\lam_6=\lam_7=0$.  One can then check
that the $\mathbold{Z}_2$
discrete symmetry is softly-broken if:
\beq
[Y^{(1)},Y]\neq 0\,,\quad {\rm and}\quad
  Z^{(1)}_{a\bbar}\propto\delta_{a\bbar} \,.
\eeq

Second, if $\lambda_5=0$ in a basis where $m_{12}^2=\lam_6=\lam_7=0$,
then the Higgs potential respects a Peccei-Quinn global U(1) symmetry.
That is, $H_1$ and $H_2$ number are individually conserved.
To find invariant conditions that must be satisfied when
this symmetry is realized, it is convenient to
work in the eigenbasis of $Y$.\footnote{If the eigenvalues of
$Y$ are degenerate, then one can work in in the eigenbasis of
$Z^{(1)}$, assuming that its eigenvalues are non-degenerate.  In this
case, the arguments above still apply with the obvious modifications.
These considerations are not applicable in
the case where \textit{both} $Y$ and $Z^{(1)}$ have degenerate eigenvalues,
which corresponds to the exceptional region of the Higgs potential
parameter space [as defined in \sect{sec:discreteA}].}
We assume that its eigenvalues are
non-degenerate so that they label $H_1$ and $H_2$.
$Z$ will conserve $H_i$ number in this basis if every
component of $Z$ has equal
number of incoming and outgoing $H_1$ lines and $H_2$ lines.
In this case, $Z$ with its incoming lines
weighted by their $Y$ eigenvalues will be equal
to $Z$ with its outgoing lines weighted by their
$Y$ eigenvalues.  This difference corresponds to the
following ``commutator-like'' fourth rank tensor:
\beq
X_{a\bbar c\dbar}\equiv Y_{a\ebar}Y_{c\fbar}Z_{e\bbar f\dbar}
-Y_{e\bbar}Y_{f\dbar}Z_{a\ebar c\fbar}\,.
\eeq

We can see that $X_{a \bbar c \dbar} = 0$ corresponds to
the Peccei-Quinn symmetry as follows.
Hermiticity properties of $Y$ and $Z$ imply that:
\beq
X_{a\bbar c\dbar}=-(X_{b\abar d\cbar})^*\,.
\eeq
Suppose that $\lam_1\neq\lam_2$ in a basis where $\lam_7=-\lam_6$.  If
$[Z^{(1)},Y]=0$ then $Y_{12}=0$.  If  $\lam_1=\lam_2$ and $\lam_7=-\lam_6$
then one can transform to a new basis in which $Y_{12}=0$.  In either case,
with $\lam_7=-\lam_6$ and $Y_{12}=0$, we find
\beq
X_{a\bbar c\dbar}\equiv (Y_{11}-Y_{22})x_{a\bbar c\dbar}\,,
\eeq
where
\beqa \label{xnum}
&& x_{1111}=0\,,\qquad\qquad\qquad\,\,\,\,\phantom{x_{2222}}
x_{2222}=0\,,\nonumber\\
&& x_{1122}=x_{2211}=0\,,\qquad\qquad\qquad\!
x_{1221}=x_{2112}=0\,,\nonumber \\
&& x_{1212}=(Y_{11}+Y_{22})\lam_5\,,\qquad\qquad
x_{2121}=-(Y_{11}+Y_{22})\lam_5^\ast\,,\nonumber\\
&& x_{1112}=x_{1211}=Y_{11}\lam_6\,,\qquad\qquad
x_{1121}=x_{2111}=-Y_{11}\lam_6^\ast\,,\nonumber \\
&& x_{2212}=x_{1222}=Y_{22}\lam_7\,,\qquad\qquad
x_{2221}=x_{2122}=-Y_{22}\lam_7^\ast\,.
\eeqa
Thus, if $Y_{11}\neq Y_{22}$, then
the additional conditions $X_{a\bbar c\dbar}=0$ guarantee that a
basis exists in which $m_{12}^2=\lam_5=\lam_6=\lam_7=0$.

If $Y_{a\bbar}\propto\delta_{a\bbar}$, then $X_{a\bbar c\dbar}=0$
automatically, and we consider instead:
\beq
\widetilde X_{a\bbar c\dbar}\equiv Z^{(1)}_{a\ebar}Z^{(1)}_{c\fbar}
Z_{e\bbar f\dbar}
-Z^{(1)}_{e\cbar}Z^{(1)}_{f\dbar}Z_{a\ebar c\fbar}\,.
\eeq
The analysis is nearly identical as for $X$ above with the
replacements $Y_{11}\to\lam_1+\lam_4$ and $Y_{22}\to\lam_2+\lam_4$.
Assuming $\lam_1\neq\lam_2$ in a basis where $\lam_7\neq -\lam_6$,
the conditions $\widetilde X_{a\bbar c\dbar}=0$ guarantee that a
basis exists in which $m_{12}^2=\lam_5=\lam_6=\lam_7=0$.
Finally, at the
exceptional point of parameter space where
$Y_{11}=Y_{22}$, $Y_{12}=0$, $\lam_1=\lam_2$ and $\lam_7=-\lam_6$, we
know it is possible to transform to a basis where $\lam_6=\lam_7=0$.
However, in general $\lam_5\neq 0$, and we have found no
additional covariant condition that requires $\lam_5=0$ in the same
basis.

\subsection{Explicit and spontaneous CP-violation}
\label{sec:discreteC}

The full power of the U(2)-invariants emerges when one studies
CP-violating theories.  Thus, the question naturally arises: what are
the invariant conditions that determine whether the theory is
CP-conserving or CP-violating.  These conditions are the analogs of
the Jarlskog invariant conditions for CP-violation~\cite{jarlskog}
in the mixing of the quark generations.
We first consider conditions such that the Higgs potential
explicitly conserves CP, independently of the value of the vacuum
expectation value.  In the generic basis, the Higgs potential
preserves the CP symmetry if there exists a U(2) transformation such
that the resulting transformed parameters of the Higgs scalar
potential are real,
\ie, $Y_{a\bbar}=(Y_{a \bbar})^\ast$ and $Z_{a\bbar c\dbar}=
(Z_{a\bbar c\dbar})^\ast$.  Using \eqs{ysym}{zsym2}, these conditions
are equivalent to: $Y_{a\bbar}=Y_{b\abar}$ and $Z_{a\bbar c\dbar}=
Z_{b\abar d\cbar}$.  However, it may be difficult in general to find
the basis where all the Higgs potential parameters are real or prove
that such a basis does not exist.  An alternative strategy is to find
U(2)-invariant quantities constructed from $Y_{a\bbar}$ and $Z_{a\bbar
c\dbar}$ that are complex in a CP-violating theory.
In an explicitly CP-conserving theory, the existence of the real basis
implies that any such invariant is real.  Hence, the non-vanishing of the
imaginary part of a potentially complex invariant would provide a
test for explicit CP-violation.

We anticipate two invariants---one involving only factors
of $Z$ tensors and one involving $Y$ and $Z$.  To see this,
we first transform to a basis where
$\lambda_6  = - \lambda_7$ [see \sect{sec:discreteA}].
The relative phase between $\Phi_1$ and
$\Phi_2$ can be chosen such that $\lambda_6$, $\lambda_7$ are both real.
Thus, assuming that $\lambda_6\neq 0$,
there remain two independent phases: ${\rm arg}(m^2_{12})$ and
${\rm arg}(\lambda_5)$.
\begin{figure}[t!]
\unitlength.5mm
\SetScale{1.418}
\begin{boldmath}
\begin{center}
\begin{picture}(100,95)(20,-40)
\ArrowLine(0,0)(40,40)
\Line(40,40)(50,50)
\ArrowLine(50,-50)(0,0)
\ArrowLine(50,50)(100,0)
\ArrowLine(100,0)(50,-50)
\ArrowArc(43,22)(15,135,495)
\Text(0,0)[c]{$\bullet$}
\Text(50,50)[c]{$\bullet$}
\Text(50,-50)[c]{$\bullet$}
\Text(100,0)[c]{$\bullet$}
\Text(33,33)[c]{$\bullet$}
\Text(5,-20)[c]{$\bullet$}
\ArrowArc(50,0)(50,0,90)
\ArrowArc(50,0)(50,90,180)
\CArc(50,0)(50,180,205)
\ArrowArc(50,0)(50,205,270)
\ArrowArc(50,0)(50,270,360)
\ArrowArc(-8,-28)(15,30,390)
\Text(50,-70)[c]{(a)}
\end{picture}
\qquad\qquad
\begin{picture}(100,95)(-20,-40)
\Text(50,-70)[c]{(b)}
\Text(0,0)[c]{$\bullet$}
\Text(100,0)[c]{$\bullet$}
\Text(70,0)[c]{$\bullet$}
\ArrowArcn(50,0)(50,180,0)
\ArrowLine(0,0)(30,0)
\ArrowLine(30,0)(70,0)
\Line(70,0)(100,0)
\ArrowArc(70,-15)(15,-270,90)
\Text(30,0)[c]{$\times$}
\Text(30,-8)[c]{$Y$}
\ArrowArcn(50,0)(50,360,180)
\ArrowArc(50,-43)(66,42,138)
\end{picture}
\end{center}
\end{boldmath}
\vspace{10mm}
\capt{Diagrams corresponding to the potentially complex invariants of
\eq{Z6Z3Y}.}
\label{fig:CPinvar}
\end{figure}
We have discovered two potentially complex invariants:
\beq
Z_{a \bar{b} c \bar{d} } Z^{(2)}_{b \bar{f}} Z^{(2)}_{d \bar{h}}
Z_{f\bar{a} j \bar{k}} Z_{k \bar{j} m \bar{n}} Z_{n \bar{m} h \bar{c}}
~~, ~~~~~~~~
Z_{a \bar{b} c \bar{d} } Y_{d \bar{e} }
Z^{(2)}_{e \bar{g}} Z_{g \bar{c} b \bar{a} }.
\label{Z6Z3Y}
\eeq
Using the diagrammatic rules introduced in \sect{sec:three}
[see Fig.~1], the invariants of eq. (\ref{Z6Z3Y})
are depicted in Fig.~3.  Applying the CP
transformation to the above diagrams changes the direction of the lines.
Hence, an invariant can be complex if its diagram looks different
when the arrows are reversed.  This requirement is satisfied for the
above diagrams.  As a result, one can easily identify
manifestly real invariants, but there is no guarantee that
an invariant corresponding to a CP-asymmetric diagram is complex.

Although we have succeeded in finding two potentially complex invariants,
it does not necessarily follow that the Higgs potential is
CP-invariant if the two invariants of \eq{Z6Z3Y} are real.  The
necessary and sufficient conditions for an
explicitly CP-invariant 2HDM potential (in terms of U(2)-invariants)
have been obtained in \Ref{cpbasis}, where it is shown that in the
$\lambda_7=-\lambda_6$ basis, both invariants of \eq{Z6Z3Y} vanish if
either $\lambda_6=0$ or if $\lambda_1=\lambda_2$.  In each of these two
cases, one must seek out a new invariant.  In \Ref{cpbasis}, the two
new invariants are found, and a proof is given showing
that the reality of all four (potentially complex)
invariants is both necessary and sufficient for an
explicitly CP-conserving Higgs potential.

If the Higgs potential is explicitly CP-conserving, it is still possible
that the Higgs vacuum does \textit{not}
respect the CP symmetry.  This is the case of spontaneously broken CP
invariance~\cite{Lee:1973iz}.
In particular, the four potentially complex invariants
noted above are real, and the so-called ``real basis'' exists in
which all the Higgs potential parameters are real.  In this basis
the Higgs vacuum expectation value
has a phase determined by the real
Higgs potential parameters
[see \eq{cpx}].
The existence of this complex phase is not necessarily
a signal for spontaneous
CP-violation.  One must prove that this phase cannot be removed
by a further change of basis (subject to the condition that the Higgs
potential parameters remain real).  Appendix~\ref{app:B} provides an example
(first discussed in \Ref{rebelo}) of a CP-conserving 2HDM in which the
complex phase of the vacuum expectation value appears in the ``real basis.''
In \Ref{cpbasis}, the explicit transformation between the two real bases
that removes the complex phase is exhibited.

Thus, it would be useful to find a set of invariant conditions
to establish the existence or non-existence of spontaneous CP-violation.
Having first proved that the Higgs potential explicitly conserves CP
(using the invariant conditions of \Ref{cpbasis}), one can prove or
rule out the existence of spontaneous CP-violation by employing
three invariants first obtained in \Ref{lavoura}.\footnote{We
prefer to express the CP-invariants in terms of the $Y_{a\bbar}$,
$Z_{a\bbar c\dbar}$ and $v_a$, following the work of \Ref{silva}.
However, only two of the three invariants were
explicitly given in this reference.}
We prove below that the
Higgs sector\footnote{In this context, the Higgs sector includes
the Higgs Lagrangian and its coupling to the gauge bosons via the
covariant derivative in the Higgs kinetic energy term.
If the Higgs-fermion Yukawa interactions are also included, then
additional CP-odd invariants based on invariants
that involve both Higgs potential parameters and the Higgs-fermion
Yukawa coupling matrices must be considered~\cite{silva}.}
conserves CP if and only if $I_1$,
$I_2$ and $I_3$ are real, where
\beqa
-\half v^2 I_1 &\equiv & \Tr~(VYZ^{(1)})\,, \label{cpxinvariants1}\\
\quarter v^4
I_2&\equiv& (VY)_{b\abar}(VY)_{d\cbar}Z_{a\bbar c\dbar}\,,
 \label{cpxinvariants2}\\
\quarter v^2 I_3 &\equiv & (VZ^{(1)})_{b\abar}V_{d\cbar}\left[\quarter v^2
Z^{(1)}_{c\ebar}Z_{a\bbar e\dbar}
+Y_{a\dbar} Z^{(1)}_{c\bbar}\right]\,.
 \label{cpxinvariants3}
\eeqa
The factors of $v^2$ have
been introduced in \eqst{cpxinvariants1}{cpxinvariants3}
so that the $I_i$ are dimensionless.
Note that $I_1$, $I_2$ and $I_3$ are invariants with respect
to the full U(2) Higgs flavor symmetry.

We may employ the scalar potential minimum conditions
[\eq{potmingeneric}] to eliminate $Y$ in the expressions for $I_1$,
$I_2$ and $I_3$:
\beqa
I_1 &\equiv &  (Z^{(1)}V)_{b\abar} V_{d\cbar}
Z_{a\bbar c\dbar}\,, \label{cpxinvariants1p}\\
I_2&\equiv& V_{a\bbar} V_{d\cbar}V_{h\gbar}V_{r \bar p}
Z_{b\ebar g\bar h}Z_{c\fbar p \bar r} Z_{e\abar f\dbar} \,,
 \label{cpxinvariants2p} \\
I_3 &\equiv & (VZ^{(1)})_{b\abar}V_{d\cbar}
\left[Z^{(1)}_{c\ebar}Z_{a\bbar e\dbar}
-2 Z^{(1)}_{c\dbar}V_{f\ebar}Z_{a\bbar e\fbar}\right]\,.
 \label{cpxinvariants3p}
\eeqa
The invariants above are most easily evaluated in the Higgs basis
[\eq{vwhiggs}].  Using \eq{hbasisinv}, we end up with:
\beq \label{cpoddinv}
\Im~I_1=\Im[Z_6 Z_7^\ast]\,,\qquad
\Im~I_2=\Im[Z_5^*Z_6^2]\,,\qquad
\Im~I_3=\Im[Z_5^*(Z_6+Z_7)^2]\,.
\eeq

The necessary and sufficient conditions for a CP-invariant scalar
Higgs potential and a CP-conserving Higgs vacuum are: $\Im~I_1=\Im~I_2=
\Im~I_3=0$.  In fact, there is at most two independent
relative phases among $I_1$, $I_2$ and $I_3$.  However, there are
cases where two of the three invariants are real and only one has an
imaginary part:\footnote{If $Z_7=0$, then $\Im~I_1=\Im~(I_2-I_3)=0$.}
\beqa
\Im~I_1&=&\Im~I_2=0\,\quad {\rm if}\quad Z_6=0\,, \label{cpxcases1} \\
\Im~I_1&=&\Im~I_3=0\,\quad {\rm if}\quad Z_7=-Z_6\,, \label{cpxcases2} \\
\Im~I_2&=&\Im~I_3=0\,\quad {\rm if}\quad Z_5=0\,, \label{cpxcases3}
\eeqa
which shows that one must check all three invariants before
determining whether the Higgs sector is CP-invariant.\footnote{If
$Y_{a\bbar}=0$ then $I_1=I_2=0$.  More generally, if
$Y_{a\bbar}\propto \delta_{a\bbar}$, then $Y_3=0$.  Consequently,
\eq{potmininv2} implies that $Z_6=0$, and $I_3$ is the only
potentially complex invariant.}

If $I_1$, $I_2$ and $I_3$ are real one can
always perform a phase rotation $H_2\to e^{i\psi}H_2$ such that
the resulting Higgs basis couplings are all real for some choice
of $\psi$.  [Note that the Higgs basis squared mass parameter,
$M^2_{12}$ is then
automatically real by virtue of \eq{hbasismincond}.]  The Higgs sector
is then manifestly CP-conserving.
Conversely, if the Higgs sector is CP-conserving, then
some basis must exist in which the Higgs potential
parameters and the Higgs field vacuum expectation values are
simultaneously real.  Thus, the \textit{invariant} quantities
$I_1$, $I_2$ and $I_3$ must be real.

\section{CP-Invariant 2HDM bosonic couplings in terms
of invariant parameters}
\label{sec:four}

The phenomenology of the two-Higgs doublet model depends in detail
on the various couplings of the Higgs bosons to gauge bosons,
Higgs bosons and fermions~\cite{hhg}.  We assume  that the Higgs sector
is CP-conserving, so we can work in a
basis in which the vacuum expectation values are both real and
non-negative, and all the parameters of the
scalar potential are real.  In this case, it is sufficient to
consider SO(2) rotations among different basis choices; hence,
there is no need to distinguish unbarred and barred
indices, and we drop the bars in what follows.
Moreover, within this set of ``real'' basis choices, the Higgs basis is
unique.  Thus, \eq{hbasisinv} implies
that the SO(2)-invariants, $Y_i$ and $Z_i$, coincide with the
Higgs potential parameters in the Higgs basis
($M_{11}^2$, $M_{22}^2$, $M_{12}^2$ and the
$\Lambda_i$, introduced in Appendix \ref{sec:two}).

Since CP is conserved by assumption,
there is no mixing between the CP-even Higgs bosons,
$\hl$ and $\hh$ and the CP-odd Higgs boson, $\ha$.  In an
arbitrary real basis, we define the angle $\alpha$ such that
\beqa
\hl &=&-(\sqrt{2}~{\rm Re\,}\Phi_1^0-v_1)\sa+
(\sqrt{2}~{\rm Re\,}\Phi_2^0-v_2)\ca\,,\nonumber\\
\hh &=&\phm(\sqrt{2}~{\rm Re\,}\Phi_1^0-v_1)\ca+
(\sqrt{2~}{\rm Re\,}\Phi_2^0-v_2)\sa\,.
\label{scalareigenstates}
\eeqa
For a basis-independent description, we define the unit vector:
\beq \label{ndef}
\widehat n=\left(
\begin{array}{c} \ca\\ \sa\end{array}\right)\,.
\eeq
Then, the two basis-independent quantities associated with the
CP-even mass eigenstates are:
\beq \label{hbasiscbma}
\widehat n_a \widehat v_a =\cosbma\equiv\cbma\,,\qquad\qquad
\epsilon_{ab} \widehat n_a \widehat v_b =\sinbma\equiv\sbma\,.
\eeq
Thus, the angle $\beta-\alpha$ represents the direction of the CP-even
Higgs mass eigenstates relative to the Higgs basis.  That is,
\beqa
\hl&=&\varphi_1^0\sinbma+\varphi_2^0\cosbma\,, \label{hbasis}\\
\hh&=&\varphi_1^0\cosbma-\varphi_2^0\sinbma\,, \label{Hbasis}
\eeqa
where $\varphi_1^0$ and $\varphi_2^0$ are defined in \eq{abbasis}.
Henceforth, we
will continue to use the notation $\sbma$ and $\cbma$ for the
basis-independent quantities [\eq{hbasiscbma}]
despite the fact that $\beta$ and
$\alpha$ are separately basis-dependent.

In the Higgs basis, the CP-even squared mass matrix takes on a rather
simple form:
\beq
\calm^2 = \left(
  \begin{array}{cc}  \Lama v^2& \Lamf v^2\\
                     \Lamf v^2& \quad\mha^2+\Lame v^2\end{array}\right)\,,
\label{massmhh}
\eeq
where
\beq \label{massmha}
\mha^2=\mhpm^2-\half v^2(\Lame-\Lamd)\,,
\eeq
and $\mhpm^2$ is given by \eq{chhiggsmass}.
After diagonalization of $\calm^2$, one determines the
CP-even Higgs squared-masses:
\beq
m^2_{\hh,\hl}=\half\left[\mha^2+v^2(\Lama+\Lame)\pm\sqrt{
[\mha^2+(\Lame-\Lama)v^2]^2+4\Lamf^2 v^4}\right]\,,
\eeq
and the angle $\beta-\alpha$ can be determined from the following results:
\beqa \label{exactbma}
\tan\,[2(\beta-\alpha)]&=&{2\Lamf v^2\over \mha^2+(\Lame-\Lama) v^2}\,,\\
\sin\,[2(\beta-\alpha)]&=&{-2\Lamf v^2\over\mhh^2-\mhl^2}\,.
\eeqa
From these expressions one can derive numerous relations among the
squared-masses, invariant coupling parameters $\Lama$, $\Lame$ and
$\Lamf$, and $\beta-\alpha$~\cite{boudjema,ghdecoupling}.

The Higgs couplings to gauge bosons follow from gauge invariance.
The properties of the three-point and
four-point Higgs boson-vector boson couplings are conveniently summarized
by listing the couplings that are proportional
to either $\sin(\beta-\alpha)$ or $\cos(\beta-\alpha)$, and the couplings
that are independent of $\beta-\alpha$~\cite{hhg}:
\beq
\renewcommand{\arraycolsep}{1cm}
\let\us=\underline
\begin{array}{lll}
\us{\cos(\beta-\alpha)}&  \us{\sin(\beta-\alpha)} &
\us{\hbox{\rm{angle-independent}}} \\ [3pt]
\noalign{\vskip3pt}
       \hh W^+W^-&        \hl W^+W^- &  \qquad\longdash   \\
       \hh ZZ&            \hl ZZ  & \qquad\longdash  \\
       Z\ha\hl&          Z\ha\hh  & ZH^+H^-\,,\,\,\gamma H^+H^-\\
       W^\pm H^\mp\hl&  W^\pm H^\mp\hh & W^\pm H^\mp\ha \\
       ZW^\pm H^\mp\hl&  ZW^\pm H^\mp\hh & ZW^\pm H^\mp\ha \\
    \gamma W^\pm H^\mp\hl&  \gamma W^\pm H^\mp\hh & \gamma W^\pm H^\mp\ha \\
   \quad\longdash    &\quad\longdash  & VV\phi\phi\,,\,VV\ha\ha\,,\,VV H^+H^-
\end{array}
\label{littletable}
\eeq
where $\phi=\hl$ or $\hh$ and $VV=W^+W^-$, $ZZ$, $Z\gamma$ or
$\gamma\gamma$.  Indeed, the Higgs boson-vector boson couplings are
basis-independent.

The three-point and four-point Higgs self-couplings are more
complicated.  Nevertheless, it is clear that one can express these
couplings in terms of $\sbma$, $\cbma$ and the invariant coupling
parameters, $\Lambda_i$.
The simplest way to obtain the Higgs self-couplings is to
work in the Higgs basis, using \eqpair{abbasis}{hbasis} and (\ref{Hbasis}).
For example,
\beqa
g\ls{\hl\hl\hl} &=& {-3v}\bigl[
    \Lama\sbma^3+\Lamcde\sbma\cbmaii+3\Lamf\cbma\sbmaii
    +\Lamg c^3_{\beta-\alpha}\bigr]\,, \\[4pt]
g_{\hl\hl\hl\hl}&=& -3\bigl[\Lama\sbma^4+\Lamb\cbma^4+2\Lamcde\cbma^2\sbma^2
             +4\Lamf\cbma\sbma^3+4\Lamg\cbma^3\sbma\bigr]
\,, \label{invfourhiggs1}
\eeqa
where $\Lamcde\equiv\Lamc+\Lamd+\Lame$.  A complete list of the
trilinear and quartic Higgs couplings in terms of the invariant
coupling parameters and $\beta-\alpha$ can be found in
\Ref{ghdecoupling}.\footnote{To make contact with the notation
of \Ref{ghdecoupling}, we define:
$\lam\equiv\Lam_1$, $\lamV\equiv\Lam_2$,
$\lamT\equiv\Lam_3+\Lam_4-\Lam_5$, $\lamF\equiv \Lam_5-\Lam_4$,
$\lamA\equiv\Lam_1-\Lam_5$, $\lamhat\equiv -\Lam_6$ and
$\lamU\equiv -\Lam_7$.}

One can generalize the analysis of this section by relaxing
the requirement that the Higgs vacuum expectation values are
positive.  That is, we consider the set of all basis choices related
by O(2) rotations.  In this case, two possible Higgs bases exist,
since one may perform the transformation $H_1\to H_1$, $H_2\to -H_2$.
Under this transformation, $M_{12}^2$, $\Lambda_6$,
$\Lambda_7$ and $\beta-\alpha$ all change sign (whereas all other
Higgs potential parameters are left unchanged).  This is not
surprising since all the SO(2)-invariants that change sign
involve an odd number of
$\epsilon_{ab}$ symbols in their definitions.  We shall use the term
\textit{pseudo-invariant} to refer to an SO(2)-invariant that changes
sign under an O(2) transformation with determinant equal to $-1$.
In addition, we note that \eqs{scalareigenstates}{higgsbasis} imply that the
physical Higgs fields are related to the Higgs basis fields by:
\beqa
\hl &=&(\sqrt{2}~{\rm Re\,}H_1^0-v)~\sbma+
\sqrt{2}~{\rm Re\,}H_2^0\,\cbma\,,\nonumber\\
\hh &=&(\sqrt{2}~{\rm Re\,}H_1^0-v)~\cbma-
\sqrt{2~}{\rm Re\,}H_2^0\,\sbma\,,
\label{hscalareigenstates}
\eeqa
$H^\pm=H_2^\pm$ and $\ha=\sqrt{2}~{\rm Im}~H_2^0$.  Thus, under
$H_1\to H_1$, $H_2\to -H_2$, the fields $\hl$, $H^\pm$ and $A$ change
sign and $\hh$ is unchanged.  One can now check that
\eqst{massmha}{invfourhiggs1} are indeed invariant with respect to O(2)
transformations, so that physical results do not depend on the choice
of basis.

\section{Yukawa couplings of the 2HDM in terms
of invariant parameters}
\label{sec:five}

The Higgs couplings to fermions are model dependent.
The most general structure for the Higgs-fermion Yukawa couplings,
often referred to as the type-III model \cite{typeiii}, is given in the
generic basis by:
\beq \label{ymodeliii}
\!\!\!\!\!\!\!\!
-\call_{\rm Y}=\anti \qlo \wtil\Phi_1\eiuo  \uro +\anti Q_L^0\Phi_1\eido \dro
+ \anti \qlo \wtil\Phi_2\eiiuo \uro +\anti \qlo \Phi_2\eiido \dro
+{\rm h.c.}\,,
\eeq
where $\Phi_{1,2}$ are the Higgs doublets, $\wtil\Phi_i\equiv
i\sigma_2 \Phi^*_i$,
$\qlo $ is the weak isospin quark doublet,
and $\uro$, $\dro$ are weak isospin quark singlets.
[The right and left-handed fermion fields are defined as usual:
$\psi_{R,L}\equiv P_{R,L}\psi$, where $P_{R,L}\equiv \half(1\pm\gamma_5)$.]
Here, $\qlo $, $\uro $,
$\dro $ denote the interaction basis states, which
are vectors in the quark
flavor space, and $\eiuo,\eiiuo,\eido,\eiido$ are matrices in quark
flavor space.  Clearly, these four matrices are basis-dependent
quantities.
We have omitted the leptonic couplings in \eq{ymodeliii};
these follow the same pattern as the down-type quark couplings.

In some models, not all the terms in \eq{ymodeliii} are present at
tree-level~\cite{hallwise}.
In a type-I model, there exist a basis where
$\eiiuo=\eiido=0$.\footnote{A type-I model can also be defined as a
model in which $\eiuo=\eido=0$ in some basis.  Clearly, the two definitions
are equivalent, since the difference in the two conditions is simply an
interchange of $\Phi_1$ and $\Phi_2$ which can be viewed as a change
of basis.}  In a type-II model, $\eiuo=\eiido=0$.  The vanishing of
certain Higgs-fermion couplings at tree-level can be enforced by a
discrete symmetry.  For example, if Lagrangian is invariant
under $\Phi_1\to +\Phi_1$, $\Phi_2\to -\Phi_2$ in some basis, then by
demanding additional discrete symmetries of a similar type for the quark
fields, one can preserve either the type-I or type-II
Higgs-fermion couplings (depending on the precise choice of
discrete symmetries) while eliminating the other possible terms in
\eq{ymodeliii}.  Another well-known example is the MSSM Higgs sector,
which exhibits a type-II Higgs-fermion coupling pattern that is enforced
by supersymmetry.

For type-I and type-II Higgs-fermion couplings, there is a natural basis
choice that is imposed, either by the discrete symmetry or
supersymmetry.  In this case, $\tanb$ becomes a meaningful parameter.
However, in the more general type-III model, there is no distinguished
basis, and once again $\tanb$ is meaningless.  To demonstrate this fact
more explicitly, we now proceed to write out the type-III Higgs-quark
interactions in a basis independent manner.
Our strategy is to rewrite the Higgs-fermion interaction
[\eq{ymodeliii}] in the Higgs
basis and identify the quark masses.
As in \sect{sec:four}, we assume that the Higgs potential is
CP-conserving so that we can work in a class of bases
related by SO(2) transformations.  Thus, the basis-independent
quantities considered in this section are invariant with respect to
SO(2).\footnote{As noted at the end of \sect{sec:four}, one can extend
the class of basis transformations to O(2).  In this case,
one should distinguish between true O(2)-invariants and pseudo-invariants
that change sign under an O(2) transformation with determinant
$-1$.  In particular, $\beta-\alpha$ and $\rho^Q$ [\eq{kapparho}] are
pseudo-invariants.  However as previously noted, the
physical Higgs fields $H^\pm$, $\ha$ and $\hl$ all change sign
under the transformation  $H_1\to H_1$, $H_2\to -H_2$ of Higgs basis
fields.  Therefore, the physical
Higgs-fermion couplings given in \eq{modeliiihqq} are invariant with
respect to the full O(2) group of basis transformations.}
However, we shall allow the Higgs-fermion couplings to be
complex (and hence CP-violating).  Using \eq{higgsbasis},
\beq \label{ymodeliiihbasis}
\!\!\!\!\!\!\!\!
-\call_Y=\anti \qlo \wtil H_1 \kpuo  \uro +\anti Q_L^0 H_1 \kpdo \dro
+ \anti \qlo \wtil H_2 \rhuo \uro +\anti \qlo  H_2 \rhdo \dro
+{\rm h.c.}\,,
\eeq
where $\wtil H_i\equiv i\sigma_2 H_i^*$ and
\beqa
\kappa^{Q,0}&\equiv& \eta^{Q,0}_1\cb+\eta^{Q,0}_2\sb\,,\label{kappadef}\\
\rho^{Q,0}&\equiv& -\eta^{Q,0}_1\sb+\eta^{Q,0}_2\cb\,, \label{rhodef}
\eeqa
for $Q=U$ or $D$. It is easy to see that $\kappa^{Q,0}$ and
$\rho^{Q,0}$ are basis-independent quantities.  First, we
introduce the vector (in Higgs flavor space) of matrix
quantities: $\eta^{Q,0}\equiv (\eta^{Q,0}_1,\eta^{Q,0}_2)$.
Then, \eqs{kappadef}{rhodef} are equivalent to:
\beq
\kappa^{Q,0}\equiv \widehat v\cdot\eta^{Q,0}\,,\qquad
\rho^{Q,0}\equiv \widehat w\cdot\eta^{Q,0}\,,
\eeq
where the dot products have the usual meaning:
$\widehat v\cdot\eta^{Q,0}\equiv \widehat v_a \eta^{Q,0}_a$
and $\widehat w\cdot\eta^{Q,0}\equiv \epsilon_{ab}\widehat v_a \eta^{Q,0}_b$.
Clearly, $\kappa^{Q,0}$ and $\rho^{Q,0}$ are SO(2)-invariant quantities.

The fermion mass eigenstates are related to the interaction eigenstates
by biunitary transformations:
\beqa
&& P_L U=V_L^U P_L U^0\,,\qquad P_R U=V_R^U P_R U^0\,,\nonumber \\
&& P_L D=V_L^D P_L D^0\,,\qquad P_R D=V_R^D P_R D^0\,,
\eeqa
and the Cabibbo-Kobayashi-Maskawa matrix is defined as $K\equiv V_L^U
V_L^{D\,\dagger}$.  It is also convenient to define ``rotated''
coupling matrices:
\beqa \label{rotyuks}
&&\eiui\equiv V_L^U \eiuoi V_R^{U\,\dagger}\,,\qquad\,
\kpu\equiv V_L^U \kpuo V_R^{U\,\dagger}\,,\qquad\,\,
\rhu\equiv V_L^U \rhuo V_R^{U\,\dagger}\,,\\
&&\eidi\equiv V_L^D \eidoi V_R^{D\,\dagger}\,,\qquad
\kpd\equiv V_L^D \kpdo V_R^{D\,\dagger}\,,\qquad
\rhd\equiv V_L^D \rhdo V_R^{D\,\dagger}\,.
\eeqa
Note that $\kappa^Q$ and $\rho^Q$ ($Q=U$, $D$) are also invariants,
since if we define $\eta^{Q}\equiv (\eta^{Q}_1,\eta^{Q}_2)$, then
\beq \label{kapparho}
\kappa^{Q}\equiv \widehat v\cdot\eta^{Q}\,,\qquad
\rho^{Q}\equiv \widehat w\cdot\eta^{Q}\,.
\eeq

The quark mass terms are identified by  replacing the scalar
fields with their vacuum expectation values.  The unitary
matrices $V_L^U$, $V_L^D$, $V_R^U$ and $V_R^D$ are then chosen so that
$\kappa^D$ and $\kappa^U$ are diagonal with real non-negative entries.
The resulting quark mass matrices are then diagonal:
\beq \label{qmasses}
M_D=\frac{v}{\sqrt{2}}\kappa^D\,,\qquad\qquad
M_U=\frac{v}{\sqrt{2}}\kappa^U\,.
\eeq

In this analysis, we have assumed that the Higgs potential is
CP-conserving, in which case we may take $\widehat
v=(\cosb\,,\,\sinb)$.  By construction, the $\kappa^Q$ are real diagonal
matrices.  \Eq{kapparho} then implies that
\beq \label{imeta}
\Im~\eta_1^Q=-\tanb\, \Im~\eta_2^Q\,,
\eeq
and $\Im~\rho^Q=\Im~\eta_2^Q/\cosb$.  Similarly, $\rho^Q$
is typically non-diagonal.  Thus, in the most general case,
$\rho^D$ and $\rho^U$ are independent \textit{complex}
non-diagonal matrices.

Finally, from \eq{ymodeliiihbasis}, we obtain the Higgs-quark
couplings after making use of \eqpair{abbasis}{hbasis} and (\ref{Hbasis}).
The end result is:
\beqa \label{modeliiihqq}
\hspace{-0.3in} -\call_Y&=&
{1\over v}\,\anti D\left[\mdd\sbma+\frac{v}{\sqrt{2}}
(\rhd P_R+{\rhd}^\dagger P_L)\cbma\right]D\hl
+\frac{i}{v}\anti D\mdd\gamma\ls{5}D\go \nonumber \\[4pt]
&&+{1\over v}\,\anti D\left[\mdd\cbma-\frac{v}{\sqrt{2}}
(\rhd P_R+{\rhd}^\dagger P_L)\sbma\right]D\hh
+\frac{i}{\sqrt{2}}\anti D(\rhd P_R-{\rhd}^\dagger P_L)D
\ha\nonumber\\[4pt]
&&+{1\over v}\,\anti U\left[\mud\sbma+\frac{v}{\sqrt{2}}
(\rhu P_R+{\rhu}^\dagger P_L)\cbma\right]U\hl
-\frac{i}{v}\anti U\mdd\gamma\ls{5}U\go \nonumber \\[4pt]
&&+{1\over v}\,\anti U\left[\mud\cbma-\frac{v}{\sqrt{2}}
(\rhu P_R+{\rhu}^\dagger P_L)\sbma\right]U\hh
-\frac{i}{\sqrt{2}}\anti U(\rhu P_R-{\rhu}^\dagger P_L)U\ha \nonumber \\[4pt]
&&+\left\{\anti U\left[K\rhd P_R-{\rhu}^\dagger KP_L\right] DH^+
+\frac{\sqrt{2}}{v}\,
\anti U\left[K\mdd P_R-\mud KP_L\right] DG^+ +{\rm h.c.}\right\}\,.
\eeqa
Once again, we observe that the Higgs interactions are determined by
basis-independent quantities (in this case, the quark masses,
$\rho^U$, $\rho^D$ and $\beta-\alpha$).
Thus, \eq{modeliiihqq} exhibits both flavor-changing Higgs-mediated neutral
currents and CP-violating Higgs-fermion couplings (even though we
assumed a CP-conserving Higgs potential, or equivalently the absence
of tree-level mixing between the CP-odd $\ha$ and the CP-even $\hl$
and $\hh$).\footnote{In principle, this requires a fine-tuning of
tree-level parameters.  If $\Im~\rho^Q\neq 0$, then one-loop effects
will generate infinite corrections to CP-violating parameters of the
Higgs potential (which are eliminated by renormalization of the
Higgs potential parameters).  The more general treatment in the case
of a tree-level CP-violating Higgs potential will be addressed in a
separate publication.}

For simplicity, we now focus on the case of one quark/lepton generation.
Then, the Higgs-quark interaction produces the following Feynman
rules of the form $-ig_{\phi f_1 f_2}$:
\beqa  \label{modeliiicouplings}
g_{\hl q\bar q}&=&
\frac{m_q}{v}\sbma+\irtwo(S_q+i\gamma_5 P_q)\cbma\,,\\[3pt]
g_{\hh q\bar q}&=&\frac{m_q}{v}\cbma-\irtwo(S_q+i\gamma_5 P_q)
\sbma\,,\\[3pt]
g_{\ha u\bar u}&=&-\irtwo(iS_u\gamma_5-P_u) \,,\\[3pt]
g_{\ha d\bar d}&=&\irtwo(iS_d\gamma_5-P_d) \,,\\[3pt]
g_{H^+ d\bar u}&=&\half[\rho^D(1+\gamma_5)-\rho^{U\ast}(1-\gamma_5)]\,,
\eeqa
where
\beq
S_q\equiv \Re~\rho^Q\,,\qquad P_q\equiv \Im~\rho^Q\,,
\eeq
and $-ig_{H^+ d\bar u}$ corresponds to the rule in which $d$, $\bar u$ and
$H^+$ are pointed into the vertex.  As noted above, if $\Im~\rho^Q\neq 0$,
then the Higgs-fermion couplings are CP-violating.

The results of \eqs{modeliiihqq}{modeliiicouplings} apply to the most
general type-III 2HDM.
To study the Higgs-fermion interactions in type-I and type-II models, it
is useful to have a basis-independent characterization of these two
special patterns of Higgs couplings.,  This is formulated
as follows:\footnote{Note that the type-I condition can be
written as
$|\eta^U\times \eta^D|^2=0$, where $\times$ corresponds to the
cross-product.  Since the Higgs flavor space is only two-dimensional,
only one component of the cross-product exists.  Nevertheless, it is
tempting to write the type-I condition as $\eta^U\times \eta^D=0$.}
\beqa
|\eta^U|^2|\eta^D|^2-|\eta^U\cdot\eta^D|^2 &=&0\,,\qquad \hbox{\rm type-I}
\,,\label{typecond1} \\
\eta^U\cdot\eta^D &=& 0\,,\qquad \hbox{\rm type-II}\,.
\label{typecond2}
\eeqa
To verify \eqs{typecond1}{typecond2}, simply note that
these equations corresponds to
the definitions of the type-I and type-II Higgs-fermion couplings in the
special basis (\ie, for the particular value of $\beta$) in which two of
the four Higgs-quark couplings of \eq{ymodeliii} vanish.  Since
\eqs{typecond1}{typecond2} are basis-independent conditions,
they must be true if they are satisfied in one particular basis.
Moreover, in both type-I and type-II models, the $\rho^Q$ are real.
This is most easily proven in the special basis by using \eq{imeta} to
show that the $\eta^Q$ must be real.

The type-I and type-II model conditions can be enforced by applying an
appropriate discrete symmetry that distinguishes between $\Phi_1$ and
$\Phi_2$~\cite{type1,hallwise,type2}.  Consequently,
$\tanb$ is promoted to a physical parameter, and thus
can be expressed in terms of invariant quantities.  For example, in a
type-II model, $\beta$ corresponds to the basis in which
$\eta_1^U=\eta_2^D=0$.  Using \eq{kapparho}, one obtains two
different equations for $\tanb$:
\beq  \label{tanbdefs}
\tanb=\frac{-\rho^D}{\kappa^D}=\frac{\kappa^U}{\rho^U}\,.
\eeq
For these two definitions to be consistent, the following equation must
be satisfied:
\beq  \label{consistcond}
\kappa^U\kappa^D+\rho^U\rho^D=0\,.
\eeq
But, \eq{consistcond} is equivalent to the type-II condition,
$\eta^U\cdot\eta^D=0$, noted above.  Moreover, using
\eqs{qmasses}{tanbdefs}, it follows that:
\beq
\rho^D=-\frac{\sqrt{2}m_d}{v}\tanb\,,\qquad \rho^U=
\frac{\sqrt{2}m_u}{v}\cotb\,.
\eeq
Inserting this result into \eq{modeliiihqq} [or
\eq{modeliiicouplings}] yields the well-known
Feynman rules for the type-II Higgs-quark interactions.
In deriving these results, the
following trigonometric identities are particularly useful:
\beqa
-{\sin\alpha\over\cos\beta}=\sbma
-\tan\beta\,\cbma\,,\qquad
\phm{\cos\alpha\over\sin\beta}=\sbma
+\cot\beta\,\cbma\,,\\[3pt]
\phm{\cos\alpha\over\cos\beta}=\cbma
+\tan\beta\,\sbma\,,\qquad
\phm{\sin\alpha\over\sin\beta}=\cbma
-\cot\beta\,\sbma\,.
\eeqa
A similar analysis can be given for models of type-I.

The analysis above makes clear that $\tanb$ is a physical parameter in
models of type-I and type-II, but is not meaningful in models of
type-III.  Nevertheless, it does suggest a strategy in the type-III
CP-conserving case.\footnote{The $\rho^Q$ are complex
if CP-violating Higgs-fermion couplings are present, in which case the
$\tan\beta$-like parameters of \eq{othertanb} would be complex.  We
shall treat the more general CP-violating case elsewhere.}
Namely, one can introduce three $\tanb$-like parameters:
\beq \label{othertanb}
\tanb_d\equiv \frac{-\rho^D}{\kappa^D}\,,\qquad
\tanb_u\equiv  \frac{\kappa^U}{\rho^U}\,,
\eeq
and a third parameter $\tanb_e\equiv -\rho^E/\kappa^E$
corresponding to the Higgs-lepton interaction.  The meaning of these
$\tanb$-like parameters is clear.  For example, consider the case of
up-type quark couplings to Higgs bosons of a type-III model.  In the
Higgs basis, both $\anti \qlo \wtil H_1 \kpuo  \uro$ and
$\anti \qlo \wtil H_2 \rhuo \uro$ interaction terms are allowed.
But, clearly there exists some basis (\ie, some rotation by angle
$\beta_u$ from the Higgs basis) for which only one of the two
up-type quark Yukawa couplings is non-vanishing.
This defines the physical angle
$\beta_u$, which is given in \eq{othertanb}.\footnote{Actually, there
is a two-fold ambiguity in the definition of $\tanb$ corresponding to
whether one identifies the Higgs boson that does not couple to the
up-type quark as $\Phi_1$ or $\Phi_2$.  The definition given in
\eq{othertanb} corresponds to the former case.  In the latter case, one
would define $\tanb_u\equiv -\rho^U/\kappa^U$.  Said another way,
suppose there exists some basis, corresponding to a rotation by an angle
$\beta_u$ from the Higgs basis, for which one of the two up-type quark
Yukawa couplings vanishes.  Then a rotation from the Higgs basis
by $\beta_u-\pi/2$ will likewise yield one vanishing
up-type quark Yukawa coupling.}
The angles $\beta_u$, $\beta_d$ and $\beta_e$ are physical
parameters of the model, and in the general type-III model there would
be no reason for these parameters to be equal in value.  However, in a
type-II model, one would indeed find that
$\tanb_u=\tanb_d=\tanb_e=\tanb$. In
some cases, the Higgs sector is
close to type-II.  For example, the MSSM Higgs sector exhibits type-II
couplings at tree-level, but all possible Higgs-fermion
couplings appear at one-loop due to
supersymmetry-breaking effects.  In this case, the Higgs-fermion
coupling is close to type-II, with differences among
the three $\tanb$-like parameters introduced above generated by
supersymmetry-breaking effects in loops corrections.

We can illustrate the last point in a very simple model approximation.
In the MSSM at large $\tanb$ (and supersymmetric masses significantly
larger than $m_Z$), the effective Lagrangian that describes the
coupling of the Higgs bosons to the third generation quarks is given
by:
\beq \label{leff}
-\call_{\rm eff}=
h_b(\anti q_L\Phi_1) b_R
+ h_t(\anti q_L \wtil\Phi_2) t_R +\Delta h_b\,(\anti q_L \Phi_2) b_R
+{\rm h.c.}\,,
\eeq
where $\anti q_L\equiv (\anti u_L\,,\,\anti d_L)$.
The term proportional to $\Delta h_b$ is generated at one-loop
due to supersymmetry-breaking effects.\footnote{One-loop corrections
to $h_b$ and $h_t$ and an effective (supersymmetry-breaking) operator
$\Delta h_t q_L\wtil\Phi_1 t_R$ yield subdominant effects at large
$\tanb$ and will therefore be neglected here.}
The tree-level relation between $m_b$ and $h_b$ is modified~\cite{deltab}:
\beq
m_b=\frac{h_b v}{\sqrt{2}}\,\cosb(1+\Delta_b)\,,
\eeq
where $\Delta_b\equiv (\Delta h_b/h_b)\tanb$.  That is, $\Delta_b$ is
$\tanb$-enhanced, and governs the leading one-loop correction
to the physical Higgs couplings to third generation quarks.
In typical models at large $\tanb$, $\Delta_b$ can be
of order 0.1 or larger and of either sign.\footnote{Explicit
expressions for $\Delta_b$ in terms of supersymmetric masses and
parameters, and references to the original literature
can be found in \Ref{Carena:2002es}.}

In the approximation scheme above, \eq{kapparho} yields
$\kappa^U\simeq h_t\sinb$, $\rho^U\simeq h_t\cosb$, and
\beqa
\kappa^D&\simeq& h_b\cosb(1+\Delta_b)\,,\\[5pt]
\rho^D &\simeq& -h_b\sinb\left(1-\frac{\Delta_b}{\tan^2\beta}\right)\simeq
-h_b\sinb\,.
\eeqa
It follows that:
\beq
\tan\beta_d\,\simeq\,\frac{\tanb}{1+\Delta_b}\,,\qquad\quad
\tan\beta_u\simeq\tan\beta\,.
\eeq
Thus, supersymmetry-breaking loop-effects can yield observable
differences between $\tan\beta$-like parameters that are defined in
terms of basis-independent quantities.

Finally, we briefly consider the multi-generation model.  The type-III
Higgs-quark interactions have already been given in \eq{modeliiihqq}.
Consider a type-II Higgs-quark interaction, defined by the matrix
equations $\eta_1^U=\eta_2^D=0$.
From \eq{kapparho}, it follows that
the following two matrix equations must be satisfied:
\beq  \label{modeliitanbdefs}
\mathbold{\it I}\tanb=-\rho^D(\kappa^D)^{-1}=\kappa^U(\rho^U)^{-1}\,,
\eeq
where $\mathbold{\it I}$ is the identity matrix in quark flavor space.
Again, we note that these equations are consistent because
$\eta^U\cdot\eta^D=0$.  Thus, using \eq{qmasses},
\beq
\rho^D=\frac{-\sqrt{2}}{v}M_D\tanb\,,\qquad
\rho^U=\frac{\sqrt{2}}{v}M_U\tanb\,,
\eeq
where $M_D$ and $M_U$ are the {\it diagonal} quark mass matrices
(implying that $\rho^D$ and $\rho^U$ are also diagonal real matrices).
We again conclude that $\tanb$ is a meaningful parameter.
The analysis for type-I models proceeds in a similar manner.
However, the construction of $\tanb$-like parameters in the general
type-III model is far more complicated.  In particular, although
the $\kappa^Q$ are diagonal, the $\rho^Q$ are complex non-diagonal
matrices in the general case.
Thus, it does not seem very useful to define $\tanb$-like
parameters by considering the non-diagonal matrices
$\kappa^U(\rho^U)^{-1}$ and $-\rho^D(\kappa^D)^{-1}$.  Fortunately, in
most models, the third-generation Yukawa couplings dominate, and one
may define $\tanb$-like parameters based solely on the consideration of
third-generation fermion couplings.

\section{Summary}
 \label{sec:six}

The scalar Lagrangian of the two-Higgs-doublet model (2HDM)
retains the same functional form under $2\times 2$ unitary
transformations among the two Higgs doublets. It is useful to understand
the effect of these (unphysical) transformations
on the Higgs potential parameters that govern the theory.
These U(2) transformations can be thought of as
generalized rotations among
different bases of Higgs fields.  The choice of basis is of course
arbitrary.

In practice, one often chooses a basis in terms of
vectors that arise naturally in the theory.  For example, in the Higgs
basis, the vacuum expectation value $v_a$ of the neutral Higgs fields
points along the direction (in Higgs flavor space) of one of the two
Higgs doublets.  If one defines $V_{a\bbar}\equiv v_a v_b^\ast$, then
the Higgs basis corresponds to the orthonormal eigenvalues of
$V_{a\bbar}$.  In a CP-conserving theory, the CP-even Higgs squared-mass
matrix is a $2\times 2$ matrix whose orthonormal eigenvalues can be
used to define the physical mass basis.  The matrix of
squared-masses, $Y_{a\bbar}$, which appears in the Higgs Lagrangian,
can also be used to define a basis in which the off-diagonal term
$Y_{12}=0$.  When fermions are coupled to the Higgs sector, discrete
symmetries are often imposed in order to guarantee the absence of
flavor-changing neutral currents.  These discrete symmetries are
defined in a particular basis, which can be related to the structure
of the Higgs-fermion interactions.

Although certain choices of basis may be physically motivated,
any underlying assumption (\textit{e.g.}, the existence of a discrete
symmetry, CP-invariance, \textit{etc.}) must be experimentally
tested.  Thus, it is especially
useful to analyze Higgs physics of the 2HDM independently
of the choice of basis.  Starting from an arbitrary
(generic) basis, one can define invariant quantities---combinations of
the Higgs potential parameters that are scalars with respect to
U(2) transformations in the flavor space of the two Higgs doublets.
Invariant descriptions of discrete symmetries and CP-symmetry
can be used in principle to test for these symmetries.
In addition, any physical Higgs sector observable (masses and couplings)
can be expressed in terms of the U(2)-invariants.  In this paper, the
relation of these observables to invariants has been obtained under
the assumption that the Higgs scalar potential is CP-invariant.  In
this latter case, it is sufficient to consider restricted O(2)-invariant
quantities.  In a subsequent publication, we will generalize the
analysis given in \sects{sec:four}{sec:five} in order to treat the
Higgs couplings in a model with a CP-violating 2HDM scalar potential.

In the most general 2HDM, the ratio of the two vacuum expectation
values, $\tan\beta$, is an unphysical basis-dependent quantity.
Typically, $\tan\beta$ acquires meaning when the
Higgs-fermion interaction is formulated, since discrete symmetries or
supersymmetry must be imposed in order to avoid flavor changing
neutral currents in conflict with experiment.  Nevertheless, these
symmetries are typically broken symmetries, so within the effective
theory of Higgs interactions, $\tan\beta$ again loses its meaning.
In this paper, we have advocated replacing $\tan\beta$ with parameters
that are more directly physical---invariant combinations of
Higgs-fermion Yukawa couplings.  Once again, by focusing
on U(2)-invariant quantities, one is led to a powerful
and flexible formalism that is ideally suited for general phenomenological
and theoretical studies of the two-Higgs-doublet model.

\bigskip

\textbf{Note Added}
\medskip

After this work was completed, a paper of
G.C.~Branco, M.N.~Rebelo and J.I.~Silva-Marcos~\cite{Branco:2005em}
appeared that emphasizes the techniques of invariants and 
addresses some of the issues considered in this paper.

\acknowledgments

This work was supported in part by the U.S. Department of Energy
and the U.K. Particle Physics and Astronomy Research Council.
We are grateful for many illuminating discussions
with Jack Gunion.  We have also benefited from conversations with
Maria Krawczyk and Ilya Ginzburg.
H.E.H. would like to express his gratitude to
W.J. Stirling and the particle theory
group in Durham for their warm hospitality during a three month
sabbatical in January--March 2003 where this project was conceived.

\appendix

\section{Basis Choices for the two-Higgs doublet model} \label{sec:two}   

In this appendix, we
review the most general two-Higgs-doublet extension of the
Standard Model~\cite{hhg,Diaz:2002tp,ghdecoupling}.
Let $\Phi_1$ and
$\Phi_2$ denote two complex $Y=1$, SU(2)$\ls{L}$ doublet scalar fields.
The most general gauge invariant scalar potential is given
by
\beqa  \label{pot}
\mathcal{V}&=& m_{11}^2\Phi_1^\dagger\Phi_1+m_{22}^2\Phi_2^\dagger\Phi_2
-[m_{12}^2\Phi_1^\dagger\Phi_2+{\rm h.c.}]\nonumber\\[6pt]
&&\quad +\half\lambda_1(\Phi_1^\dagger\Phi_1)^2
+\half\lambda_2(\Phi_2^\dagger\Phi_2)^2
+\lambda_3(\Phi_1^\dagger\Phi_1)(\Phi_2^\dagger\Phi_2)
+\lambda_4(\Phi_1^\dagger\Phi_2)(\Phi_2^\dagger\Phi_1)
\nonumber\\[6pt]
&&\quad +\left\{\half\lambda_5(\Phi_1^\dagger\Phi_2)^2
+\big[\lambda_6(\Phi_1^\dagger\Phi_1)
+\lambda_7(\Phi_2^\dagger\Phi_2)\big]
\Phi_1^\dagger\Phi_2+{\rm h.c.}\right\}\,,
\eeqa
where $m_{11}^2$, $m_{22}^2$, and $\lam_1,\cdots,\lam_4$ are real parameters.
In general, $m_{12}^2$, $\lambda_5$,
$\lambda_6$ and $\lambda_7$ are complex.
The scalar fields will
develop non-zero vacuum expectation values if the mass matrix
$m_{ij}^2$ has at least one negative eigenvalue.
We assume that the
parameters of the scalar potential are chosen such that
the minimum of the scalar potential respects the
U(1)$\ls{\rm EM}$ gauge symmetry\cite{Ferreira:2004yd}.
Then, the scalar field
vacuum expectations values are of the form\footnote{In writing \eq{potmin},
we have used a global SU(2)$_W$ rotation to
put the non-zero vacuum expectation values
in the lower component of the doublet, and
a global hypercharge U(1) rotation to eliminate the phase
of~$v_1$.}
\beq \label{potmin}
\langle \Phi_1 \rangle={1\over\sqrt{2}} \left(
\begin{array}{c} 0\\ v_1\end{array}\right), \qquad \langle
\Phi_2\rangle=
{1\over\sqrt{2}}\left(\begin{array}{c}0\\ v_2\, e^{i\xi}
\end{array}\right)\,,
\eeq
where $v_1$ and $v_2$ are real and non-negative, $0\leq \xi <2\pi$  and
\beq \label{v246}
v^2\equiv v_1^2+v_2^2={4\mw^2\over g^2}=(246~{\rm GeV})^2\,.
\eeq
The corresponding potential minimum conditions are:
\beqa
\hspace{-1in}
m_{11}^2 \!&=&\! m_{12}^2\,e^{i\xi}\,\tb  -\half
v^2\left[\lam_1\cb^2+(\lam_3+\lam_4+\lam_5 e^{2i\xi})\sb^2
+(2\lam_6 e^{i\xi}\!+\!
\lam_6^\ast e^{-i\xi})\sb\cb+\lam_7\sb^2\tb e^{i\xi}\right]
\label{minconditionsa} \\[6pt]
\hspace{-1in}
m_{22}^2 \!&=&\! (m_{12}^2\,e^{i\xi})^\ast\,\tb^{-1} -\half v^2
\left[\lam_2\sb^2+(\lam_3+\lam_4+\lam_5^\ast e^{-2i\xi})\cb^2
+\lam_6^\ast \cb^2\tb^{-1}e^{-i\xi}\right. \nonumber \\
&&\qquad\qquad\qquad\qquad\qquad\qquad\qquad\qquad\qquad\qquad\quad\,\,\left.
+(\lam_7e^{i\xi}+2\lam_7^\ast e^{-i\xi})\sb\cb\right],
\label{minconditionsb}
\eeqa
where
\beq
\sb\equiv\sinb\equiv\frac{v_2}{v}\,,\qquad
\cb\equiv\cosb\equiv\frac{v_1}{v}\,,\qquad {\rm and}\qquad
\tb\equiv\tanb\equiv\frac{v_2}{v_1}\,,
\label{tanbdef}
\eeq
and $0\leq \beta\leq\pi/2$ (since by assumption, $v_1$, $v_2\geq 0$).
Since $m_{11}^2$ and $m_{22}^2$ are both real, the imaginary part of
either
\eq{minconditionsa} or \eq{minconditionsb} yields one independent equation:
\beq \label{cpx}
\Im(m_{12}^2 e^{i\xi})=\half v^2\left[\Im(\lam_5 e^{2i\xi})\sb\cb
+\Im(\lam_6 e^{i\xi})\cb^2+\Im(\lam_7 e^{i\xi})\sb^2\right]\,.
\eeq
\Eq{cpx} can be used to determine $\xi$.

Of the original eight scalar degrees of freedom, three Goldstone
bosons ($G^\pm$ and~$\go$) are absorbed (``eaten'') by the $W^\pm$ and
$Z$.  These states are easily identified:
\beqa
G^\pm &=& \cosb\,\Phi_1^\pm+e^{-i\xi}\,\sinb\,
\Phi_2^\pm\,,\label{gpm}\\
\go\phantom{^\pm}\!\!&=&
\sqrt{2}\left[\cosb\,{\rm Im}~\Phi_1^0+\sinb\,
{\rm Im}~(e^{-i\xi}\,\Phi_2^0)\right]\,.
\label{goldn}
\eeqa

In writing all the expressions above, we have implicitly chosen a basis
in the space of $\Phi_1$--$\Phi_2$.  We shall refer to this basis choice as
the \textit{generic} basis.  One particular basis choice is especially
useful.  This is the so-called Higgs basis of \Ref{silva} (whose significance
was also emphasized in refs.~\cite{Donoghue:1978cj,Georgi,lavoura,lavoura2})
in which only the neutral
component of one of the two Higgs doublets (\eg, $\Phi_1^0$)
possesses a vacuum expectation value.  This requirement
is not sufficient to uniquely define the Higgs basis.
In particular, there
is a set of U(1)$_{\rm Y}\times$U(1) transformations on the scalar field that
preserves the conditions: $|\vev{\Phi_1^{\prime\,0}}|=v/\sqrt{2}$ and
$\vev{\Phi_2^{\prime\,0}}=0$, where the $\Phi'_i$ are the scalar fields
in the basis where $\beta=0$~\cite{branco}.
U(1)$_{\rm Y}$ is  the global hypercharge
transformation, and is a {\it symmetry } of the Lagrangian,
unlike the SU(2) transformations in Higgs flavor space which
transform the parameters of the Higgs potential.
The second U(1) is the diagonal subgroup of the SU(2)
in Higgs flavor space,
corresponding to $\Phi'_1\to e^{i\chi}\Phi_1'$ and $\Phi'_2\to
e^{-i\chi}\Phi_1'$.
   Thus, there is a family of Higgs bases,
parameterized by $\chi$.\footnote{The range of $\chi$ can be taken to
be $0\leq\chi\leq\pi$, since the transformation $\chi\to\chi+\pi$ can
be compensated by a U(1)$_{\rm Y}$ rotation.}
Let us denote the two Higgs
doublets in the Higgs basis by $H_1$ and $H_2$.
Then, the relation
between these scalar doublet fields and the scalar fields in the
original basis is given by:
\beqa \label{higgsbasis}
e^{-i\chi}\,H_1&=&\Phi_1\cosb+e^{-i\xi}\,\Phi_2\sinb\,,\nonumber \\
e^{i\chi}\,H_2&=&-e^{i\xi}\,\Phi_1\sinb+\Phi_2\cosb\,.
\eeqa
One then obtains
\beq \label{abbasis}
e^{-i\chi}\,H_1=\left(\begin{array}{c}
G^+ \\ {1\over\sqrt{2}}\left(v+\varphi_1^0+iG^0\right)\end{array}
\right)\,,\qquad
e^{i\chi}\,H_2=\left(\begin{array}{c}
H^+ \\ {1\over\sqrt{2}}\left(\varphi_2^0+i\ha\right)\end{array}
\right)\,,
\eeq
where $\varphi_1^0$, $\varphi_2^0$ are CP-even neutral Higgs fields,
$\ha$ is a CP-odd neutral Higgs field, and $H^+$ is the physical
charged Higgs boson.  If the Higgs sector is CP-violating, then
$\varphi_1^0$, $\varphi_2^0$, and $\ha$ all mix to produce three
physical neutral Higgs states of indefinite CP.  If CP is conserved,
then $\ha$ is the physical CP-odd Higgs scalar and $\varphi_1^0$
and $\varphi_2^0$ mix to produce two physical neutral CP-even Higgs
states $\hl$ and $\hh$. We examine the CP-conserving case in more
detail in \sect{sec:four}.

In the Higgs basis, the corresponding values of $\lam_1,\cdots,\lam_7$
can be easily computed by re-expressing $\Phi_1$ and $\Phi_2$ in terms
of $H_1$ and $H_2$ [\eq{higgsbasis}] and inserting the
result into \eq{pot}.  The end result is:
\beqa  \label{pothbasis}
\mathcal{V}&=& M_{11}^2 H_1^\dagger H_1+M_{22}^2 H_2^\dagger H_2
-[M_{12}^2 H_1^\dagger H_2+{\rm h.c.}]\nonumber\\[4pt]
&&\quad\!\!\!\! +\half\Lama(H_1^\dagger H_1)^2
+\half\Lamb(H_2^\dagger H_2)^2
+\Lamc(H_1^\dagger H_1)(H_2^\dagger H_2)
+\Lamd(H_1^\dagger H_2)(H_2^\dagger H_1)
\nonumber\\[4pt]
&&\quad\!\!\!\! +\left\{\half\Lame(H_1^\dagger H_2)^2
+\big[\Lamf\,(H_1^\dagger H_1)
+\Lamg(H_2^\dagger H_2)\big]
H_1^\dagger H_2+{\rm h.c.}\right\}\,,
\eeqa
where
\beqa
M_{11}^2&=&m_{11}^2\cb^2+m_{22}^2\sb^2-\Re(m_{12}^2 e^{i\xi})\stwob\,,
\label{maa}\\
M_{22}^2&=&m_{11}^2\sb^2+m_{22}^2\cb^2+\Re(m_{12}^2 e^{i\xi})\stwob\,,
\label{mbb}\\
M_{12}^2 e^{i(\xi-2\chi)}&=&
\half(m_{11}^2-m_{22}^2)\stwob+\Re(m_{12}^2 e^{i\xi})\ctwob
+i\,\Im(m_{12}^2 e^{i\xi})\,.\label{mab}
\eeqa
and
\beqa
\!\!\!\!\!\!\!\!\!\!\!\!\!\!\Lama&=&
\lam_1\cb^4+\lam_2\sb^4+\half\lamtil\stwob^2
+2\stwob\left[\cb^2\Re(\lam_6 e^{i\xi})
+\sb^2\Re(\lam_7e^{i\xi})\right]\,,
\label{Lam1def}  \\
\!\!\!\!\!\!\!\!\!\!\!\!\!\!\Lamb &=&
\lam_1\sb^4+\lam_2\cb^4+\half\lamtil\stwob^2
-2\stwob\left[\sb^2\Re(\lam_6 e^{i\xi})
+\cb^2\Re(\lam_7e^{i\xi})\right]\,,
\label{Lam2def}      \\
\!\!\!\!\!\!\!\!\!\!\!\!\!\!\Lamc &=&
\quarter\stwob^2\left[\lam_1+\lam_2-2\lamtil\right]
+\lam_3-\stwob\ctwob\Re[(\lam_6-\lam_7)e^{i\xi}]\,,
\label{Lam3def}      \\
\!\!\!\!\!\!\!\!\!\!\!\!\!\!\Lamd &=&
\quarter\stwob^2\left[\lam_1+\lam_2-2\lamtil\right]
+\lam_4-\stwob\ctwob\Re[(\lam_6-\lam_7)e^{i\xi}]\,,
\label{Lam4def}      \\
\!\!\!\!\!\!\!\!\!\!\!\!\Lame e^{2i(\xi-2\chi)}&=&
\quarter\stwob^2\left[\lam_1+\lam_2-2\lamtil\right]+\Re(\lam_5 e^{2i\xi})
+i\ctwob\Im(\lam_5 e^{2i\xi})
\nonumber \\
&&\qquad
-\stwob\ctwob\Re[(\lam_6-\lam_7)e^{i\xi}]
-i\stwob\Im[(\lam_6-\lam_7)e^{i\xi})]\,,
\label{Lam5def}      \\
\!\!\!\!\!\!\!\!\!\!\!\!\Lamf e^{i(\xi-2\chi)}&=&
-\half\stwob\left[\lam_1\cb^2
-\lam_2\sb^2-\lamtil\ctwob-i\Im(\lam_5 e^{2i\xi})\right]\nonumber \\
&&\qquad
+\cb\cthreeb\Re(\lam_6 e^{i\xi})+\sb\sthreeb\Re(\lam_7 e^{i\xi})
+i\cb^2\Im(\lam_6 e^{i\xi})+i\sb^2\Im(\lam_7 e^{i\xi})\,,
\label{Lam6def}      \\
\!\!\!\!\!\!\!\!\!\!\!\!\Lamg e^{i(\xi-2\chi)}&=&
-\half s_{2\beta}\left[\lam_1\sb^2-\lam_2\cb^2+
\lamtil c_{2\beta}+i\Im(\lam_5 e^{2i\xi})\right]\nonumber \\
&&\qquad
+\sb\sthreeb\Re(\lam_6 e^{i\xi})+\cb\cthreeb\Re(\lam_7e^{i\xi})
+i\sb^2\Im(\lam_6 e^{i\xi})+i\cb^2\Im(\lam_7e^{i\xi})\,,
\label{Lam7def}
\eeqa
where
\beq \label{lamtildef}
\lamtil\equiv\lam_3+\lam_4+\Re(\lam_5 e^{2i\xi})\,.
\eeq
The inversion of \eqst{maa}{mab} and
\eqst{Lam1def}{Lam7def} is simply obtained by making the replacements
$M^2_{ij}\tto m^2_{ij}$, $\Lambda_i\tto\lambda_i$, $\beta\tto
-\beta$ and $\xi\tto\xi-2\chi$.
Note that the Higgs basis corresponds to the choice $\beta=0$,
independently of the value of~$\xi$.  The effect of
non-uniqueness of the Higgs
basis, which is parameterized by $\chi$, is also easily discerned.
Starting from the Higgs basis with $\chi=0$, one can transform to the
Higgs basis with arbitrary $\chi$ by the phase redefinitions
\beq \label{redefinitions}
H_1\to e^{i\chi}H_1 \quad {\rm and} \quad
H_2\to e^{-i\chi} H_2\,.
\eeq
As a result of this transformation,
the complex parameters of the scalar potential in the Higgs basis are
transformed by a phase rotation:
$M_{12}^2,\Lambda_6,\Lambda_7 \to e^{2i\chi}[M_{12}^2,\Lambda_6,\Lambda_7]$,
and $\Lambda_5\to e^{4i\chi}\Lambda_5$.

Finally, the scalar potential minimum conditions in Higgs basis are
independent of $\chi$:
\beq \label{hbasismincond}
M_{11}^2=-\half\Lama v^2\,,\qquad\qquad M_{12}^2=\half\Lamf v^2\,.
\eeq
Note that the equation for
$M_{12}^2$ is a complex equation that holds both for the real and
imaginary parts.
The value of $M_{22}^2$ is not constrained by the
scalar potential minimum conditions.
One can show that $M_{22}^2$ is
directly related to the physical charged Higgs boson squared-mass:
\beq \label{chhiggsmass}
\mhpm^2=M_{22}^2+\half v^2\Lamc\,,
\eeq
where $H^\pm=-e^{i\xi}\sinb\,\Phi_1^\pm+\cosb\,\Phi_2^\pm$
is the state orthogonal to
the charged Goldstone boson [\eq{gpm}].
The requirement that $\mhpm^2\geq 0$ places a constraint on $M_{22}^2$
and $\Lamc$ since we have assumed that electromagnetism is not
spontaneously broken.

We can now count the number of independent parameters that govern the
most general 2HDM.  After imposing the scalar potential minimum
conditions in the Higgs basis, there are twelve
parameters: $M_{11}^2$ (which determines the value of
$v=246$~GeV  [see \eq{hbasismincond}]),
$M_{22}^2$, four real couplings $\Lama,\ldots,\Lamd$
and three complex couplings $\Lame$, $\Lamf$ and $\Lamg$.  However,
the non-uniqueness of the Higgs basis implies that only the relative
phases of the these complex parameters are physical.
That is, of the twelve independent parameters that characterize the
Higgs basis, one degree of freedom can be removed by the phase
redefinitions given by \eq{redefinitions}, leaving
only eleven physical degrees of freedom~\cite{lavoura2}.

To see that this result is consistent with the parameter counting in the
generic basis, we first note that without loss of generality, we can
choose a basis in which $m_{12}^2=0$. In this case, $\cos^2 2\beta$
becomes a physical parameter as noted at the end of
\sect{sec:three}.  Moreover, we can make an additional
phase rotation $\Phi_2\to e^{i\xi}\,\Phi_2$ such that the
two neutral scalar vacuum expectation values are real.  The minimum
conditions [\eqs{minconditionsa}{minconditionsb}] fix the values of
$m_{11}^2$ and $m_{22}^2$, leaving twelve parameters: $v_1$, $v_2$,
the real couplings $\lam_1,\ldots,\lam_4$ and the complex couplings
$\lam_5, \lam_6$ and $\lam_7$.  But, in the basis where $m_{12}^2=0$,
the imaginary part of the minimum conditions
[\eq{cpx}] implies one relation among $\Im~\lam_5$, $\Im~\lam_6$ and
$\Im~\lam_7$.  Thus, we again end up with
eleven independent parameters, as expected.
We may also obtain the same result by employing
an elegant technique advocated in \Refs{santamaria}{sutter}.
In the generic basis, the scalar Higgs potential consists of
six real and four complex parameters.  If the Higgs potential
is set to zero, the scalar Lagrangian possesses a U(2) global symmetry
corresponding to arbitrary U(2) transformations in the ``flavor''
space of Higgs doublets $\{\Phi_1\,,\,\Phi_2\}$.   In particular, note
that U(2)~$\iso$~SU(2)$\times$U(1)$_{\rm Y}$, where the SU(2) is a
global flavor symmetry not to be confused with the gauged SU(2) of the
Standard Model.  When the Higgs
potential is restored, the scalar Lagrangian is no longer invariant
under the SU(2) global flavor transformations.
That is, the effect of the flavor SU(2) transformations is to transform the
Higgs potential parameters.  However, we are always free to redefine
the scalar fields by the flavor SU(2) transformation.  A general SU(2)
transformation is determined by three parameters.  Thus, three of the
fourteen Higgs potential parameters can be transformed away (or set to
zero) by an appropriate flavor SU(2) redefinition of the two scalar
doublet fields.  This leaves eleven physical degrees of freedom,
again confirming our previous counting.

\section{Invariants by explicit construction of eigenvectors}
\label{app:three}

In this Appendix, we explicitly construct eigenvectors of the
hermitian matrix $V_{a\bbar}\equiv \widehat v_a \widehat v_{\bbar}^*$
[where $\widehat v$ is the unit vector
of Higgs vacuum expectation values defined in \eq{sachavev}],
and contract them with the Higgs potential parameter tensors
($Y$ and $Z$) to obtain index-free ``objects''. By
carefully considering the phase transformations
on the eigenvectors, we will see that some of our
objects are invariants with respect to U(2), and some pick up
phases under U(2) transformations. The latter set can be
combined into the invariants of \sect{sec:three}.

Consider the two orthonormal eigenvectors of $V$:
\beq \label{vhat}
\widehat v_a \equiv e^{i\eta}\left(
\begin{array}{c} \cb\\ \sb\,e^{i\xi} \end{array}\right)
\,,\qquad\qquad
\widehat w_a \equiv -\epsilon_{ab}\,\widehat
v_\bbar^\ast=e^{-i\eta}\left(
\begin{array}{c} -\sb\,e^{-i\xi}\\ \cb\end{array}\right)\,,\
\eeq
with the inverse relation $\widehat v^\ast_{\abar}=\epsilon_{\abar\bbar}
\,w_b$.  In \eqs{sachavev}{potmin},
a global U(1)$_Y$ rotation has been employed to eliminate
the overall phase factor $e^{i\eta}$ in $\widehat v$.
However, it will sometimes
be convenient to use the more general forms given in \eq{vhat}.

Given a generic basis and the corresponding form for $\widehat v$, we
can transform to another basis by redefining the scalar fields via
$\Phi'_{a}=U_{a\bbar}\Phi_b$.
The most general U(2) transformation is given by
\beq \label{utransform}
U=e^{i\psi}
\left(\begin{array}{cc} e^{i\gamma}\cos\theta &
\quad e^{-i\zeta}\,\sin\theta  \\
 -e^{i\zeta}\,\sin\theta & \quad e^{-i\gamma}\,\cos\theta
\end{array}\right)\,,
\eeq
where $0\leq\theta\leq\pi/2$, and
$0\leq\gamma\leq\pi$
and $0\leq\zeta\,,\,\psi<2\pi$.\footnote{Note that for fixed
$\theta$, $U$ does not change when $\psi$, $\gamma$, $\zeta\to
\pi+[\psi$, $\gamma$, $\zeta$].  Thus, we have chosen to restrict $\chi$
to lie between $0$ and $\pi$ (other choices are possible).  If
$\psi=0$, then \eq{utransform} provides a parameterization of
SU(2)$/\mathbold{Z_2}$.  In this case, the full SU(2)
group manifold can be covered by
extending the range of $\gamma$ to $0\leq\gamma < 2\pi$.}
The subgroup of U(1)$_{\rm Y}$ global hypercharge rotations
corresponds to freezing the values of
$\theta$, $\gamma$ and $\zeta$, and the subgroup of the flavor SU(2)
transformations corresponds to freezing the value of $\psi$.
Starting from the generic basis, we first make a U(1)$_{\rm Y}$
transformation to set $\eta=0$ in \eq{vhat}.  Then, we may employ the
flavor SU(2) transformation given by $\psi=0$\,,\, $\gamma=\chi$\,,\,
$\zeta=\xi-\chi$ and
$\theta=\beta$ to rotate from the generic basis to the Higgs basis as
specified by \eq{higgsbasis}.  Applying this form of
$U$ to $\widehat v$ and $\widehat w$ [\eq{vhat} with $\eta=0$]
yields the corresponding results in the Higgs basis:
$\widehat v=(e^{i\chi},0)$ and $\widehat w=(0,e^{-i\chi})$.
The parameter $\chi$ reflects the non-uniqueness in the definition of
the Higgs basis, as discussed in Appendix \ref{sec:two}.

We now turn to the task of constructing
U(2)-invariant (scalar) combinations of
Higgs potential parameters.
We shall also construct objects that are invariant
with respect to the flavor SU(2) transformations, but not
necessarily invariant with respect to the full U(2) group.
These objects  are useful, but they do not correspond
directly to physical quantities.

We may combine the tensors $Y_{a\bbar}$, $Z_{a\bbar c\dbar}$,
$\widehat v_a$, $\widehat v_{\abar}^\ast$,  $\widehat w_a$ and
$\widehat w_{\abar}^\ast$ to create scalar
quantities by summing over pairs of indices (following the rule that
the summed index pairs must contain one unbarred and one barred index).
A U(2) transformation can be written as
$U\equiv e^{i\psi}\widehat U$ with $\det~\widehat U=1$.
With respect to the flavor-SU(2), both $\widehat v$ and $\widehat w$
transform covariantly:\footnote{The fact that $\widehat w$
transforms covariantly under SU(2) transformations, $\widehat U$,
is a consequence of the well-known identity among Pauli matrices:
$\sigma_2\mathbold{\vec\sigma}\sigma_2=\mathbold{\vec\sigma}\lss{\ast}$.
This identity can be used to prove the relation $\epsilon_{ab}
\widehat U^\dagger_{c\bbar} =\widehat{U}_{a\bbar}\epsilon_{bc}$.}
$\widehat v_a\to
\widehat{U}_{a\bbar}\widehat v_b$ then $\widehat w_a\to
\widehat{U}_{a\bbar} \widehat w_b$.
With respect to U(1)$_{\rm Y}$, $\widehat{v}$ inherits the transformation
law of $\Phi_a$.  Hence with respect to U(2) transformations, $\widehat v_a\to
U_{a\bbar}\widehat v_b$.  In contrast,
\beq \label{what}
\widehat w_a\to (\det U)^{-1} U_{a\bbar} \widehat w_b\,.
\eeq
That is, $\widehat v$ and $\widehat w$ transform oppositely with
respect to U(1)$_{\rm Y}$ as expected.

It is now straightforward to
construct index-free objects that are
linear in the Higgs potential parameters.
We obtain three squared-mass invariants:
\beq
Y_1 \equiv  Y_{a\bbar}\,\widehat v_\abar^\ast\, \widehat v_b\,,
\qquad\qquad
Y_2 \equiv Y_{a\bbar}\,\widehat w_\abar^\ast\, \widehat w_b\,,
\qquad\qquad
Y_3 \equiv  Y_{a\bbar}\,\widehat v_\abar^\ast\, \widehat w_b
\,,\label{yvv}
\eeq
and seven coupling invariants:
\beqa
Z_1\equiv Z_{a\bbar c\dbar}\,\widehat v_\abar^\ast\, \widehat v_b\,
\widehat v_\cbar^\ast\,\widehat v_d\,,\phantom{i}\label{zvv1}& ~~~~~~~&
Z_2 \equiv Z_{a\bbar c\dbar}\,\widehat w_\abar^\ast\, \widehat w_b\,
\widehat w_\cbar^\ast\,\widehat w_d\,,\\
Z_3 \equiv Z_{a\bbar c\dbar}\,\widehat v_\abar^\ast\, \widehat v_b\,
\widehat w_\cbar^\ast\,\widehat w_d\,,\label{zvv3}&~~~~~~~&
Z_4 \equiv
Z_{a\bbar c\dbar}\, \widehat w_\abar^\ast\, \widehat v_b\,
\widehat v_\cbar^\ast\,\widehat w_d\,,\label{zvv4} \\
&& \hspace{-1in} Z_5 \equiv
Z_{a\bbar c\dbar}\,\widehat v_\abar^\ast\, \widehat w_b\,
\widehat v_\cbar^\ast\, \widehat w_d
\,,\label{zvv5}\\
&& \hspace{-1in}
Z_6 \equiv  Z_{a\bbar c\dbar}\,\widehat v_\abar^\ast\,\widehat v_b\,
\widehat v_\cbar^\ast\, \widehat w_d\,,\label{zvv6}\\
&& \hspace{-1in} Z_7 \equiv
     Z_{a\bbar c\dbar}\,\widehat v_\abar^\ast\, \widehat w_b\,
\widehat w_\cbar^\ast\,\widehat w_d\,.\label{zvv7}
\eeqa

Noting the transformation law for $w$ [\eq{what}], it then follows
that the real quantities $Y_1$, $Y_2$, $Z_{1,2,3,4}$ are
U(2)-invariants, whereas the complex quantities $Y_3$ and $Z_{5,6,7}$
are only flavor-SU(2)-invariants.  However, from the four complex
flavor-SU(2)-invariants, we can construct seven real U(2)-invariant
quantities: the magnitudes $|Y_3|$ and $|Z_{5,6,7}|$, and three
relative phases: arg($Y_3^2 Z_5^*$), arg($Y_3 Z_6^*$) and arg($Y_3 Z_7^*$).
Thus, we have recovered the thirteen U(2)-invariant quantities
given in \eqst{syvv1}{szvv7}.  Imposing the scalar potential minimum
conditions [\eqs{potmininv1}{potmininv2}]
reduces the number of physical degrees
of freedom to eleven as expected.

Let us first evaluate \eqst{yvv}{zvv7}
in the Higgs basis where $\widehat v=(e^{i\chi},0)$ and
$\widehat w=(0,e^{-i\chi})$.
By inspection, one obtains
\beqa \label{hbasisinv}
Y_1&=& M_{11}^2\,,\qquad\,\,Y_2=M_{22}^2,\qquad\, Y_3=-M_{12}^2e^{-2i\chi}\,,
\qquad\, Z_i=\Lambda_i\quad (i=1,\ldots,4)\,, \nonumber \\
Z_5&=& \Lambda_5 e^{-4i\chi}\,,\qquad
Z_6=\Lambda_6 e^{-2i\chi}\,,\qquad
Z_7=\Lambda_7 e^{-2i\chi}\,.
\eeqa

The appearance of the $\chi$-dependent phases in the expressions for
the complex invariants in \eq{hbasisinv}
is expected.  As noted below
\eq{redefinitions}, if one transforms from the Higgs basis with
$\chi=0$ to a Higgs basis with arbitrary $\chi$, then
$M_{12}^2,\Lambda_6,\Lambda_7 \to e^{2i\chi}[M_{12}^2,\Lambda_6,\Lambda_7]$,
and $\Lambda_5\to e^{4i\chi}\Lambda_5$.  But, this is precisely what
is needed to ensure that $Y_3$ and $Z_{5,6,7}$ are invariant with
respect to the flavor-SU(2) transformation $U_D\equiv {\rm diag}(e^{i\chi},
e^{-i\chi})$. One can also evaluate \eqst{yvv}{zvv7} in the
generic basis where $\widehat v$ and $\widehat w$ are given
by \eq{vhat} with $\eta=0$.  The result of this computation
simply reproduces the results of \eqst{maa}{Lam7def}.

Using \eqs{zsym1}{zsym2}, one can show that \eqst{yvv}{zvv7} exhaust
all possible independent invariants that are linear in the Higgs
potential parameters.  For example, by inserting $\widehat
w_a=-\epsilon_{ab}\widehat v_b^*$ into \eqst{yvv}{zvv4}, one finds:
\beqa
Y_1+Y_2 &=&\Tr~Y\,,\label{morey}\\
Z_1+Z_3 &=& Z^{(2)}_{a\bbar} v_{\abar}^\ast v_b\,,\label{morez1}\\
Z_1+Z_4 &=& Z^{(1)}_{a\bbar} v_{\abar}^\ast v_b\,,\label{morez2}\\
Z_1+Z_2+2Z_3 &=&\Tr~Z^{(2)}\,,\label{morez3}\\
Z_1+Z_2+2Z_4 &=&\Tr~Z^{(1)}\,,\label{morez4}
\eeqa
which demonstrates that $Z^{(k)}_{a\bbar} v_{\abar}^\ast v_b$ and
$\Tr~Z^{(k)}$ ($k=1,2$) are not independent invariants.
Note that the latter two invariant quantities are independent of the
vacuum expectation values, and thus correspond to combinations of the
Higgs self couplings that are invariant under an arbitrary change of
basis.\footnote{That is, $\Tr~Z^{(2)}\equiv
Z_{a\abar b\bbar}=\lam_1+\lam_2+2\lam_3$ and $\Tr~Z^{(1)}\equiv
Z_{a\bbar b\abar}=\lam_1+\lam_2+2\lam_4$, with respect to any basis choice.}
Furthermore,
$\epsilon_{\abar\cbar}\,\epsilon_{bd}
\,Z_{a\bbar c\dbar}=\Tr~[Z^{(2)}-Z^{(1)}]=\lam_3-\lam_4$
is not independent from the invariants of
\eqs{morez3}{morez4}.\footnote{Once can also define
$X_{ab}\equiv\epsilon_{cd} Z_{a\cbar b\dbar}$.  However, employing
this tensor does not lead to any genuinely new invariants.  One can
easily verify that $X_{ab}=(\lam_3-\lam_4)\,\epsilon_{ab}$, which we
recognize as an SU(2)-invariant tensor.}

If we restrict our considerations to
a CP-conserving theory, where all scalar coupling
parameters may be taken real and basis changes are restricted to those
related by O(2) transformations, then the distinction between barred and
unbarred indices becomes irrelevant.  In this case, one more
independent invariant arises that is linear in the couplings:
$Z_{abab}=\lambda_1+\lambda_2+2\lambda_5$, with respect to any
real basis choice.\footnote{Note that \eqst{Lam1def}{Lam5def}
yield the relation
$\Lama+\Lamb+2~\Re(\Lame e^{2i(\xi-2\chi)})=\lam_1+\lam_2+2~\Re(\lam_5
e^{2i\xi})$.   However, one cannot extract from this result a
quantity that is invariant under the most general
flavor SU(2) and U(2) transformations.}

Likewise, one can also construct additional
invariants involving higher powers of the mass terms
$Y$ and the couplings $Z$.  For example, $2\times 2$ matrices satisfy
\beq
{\rm det}~Y
=\half\epsilon_{bd}\,\epsilon_{\abar\cbar}\,Y_{a\bbar}\,Y_{c\dbar}=
\half\left[(\Tr~Y)^2-\Tr~Y^2\right]\,.
\eeq
This yields the (invariant) determinant.  One can also construct invariants
that depend on inverse powers of $Y$ [since
$Y^{-1}$ transforms the same way as $Y$ under U(2)].
However, according to the Cayley-Hamilton theorem,
all matrices satisfy their characteristic equations~\cite{linalg}.
Consequently, $({\rm det}~Y)[Y^{-1}]_{a\bar b}=
({\rm Tr}~Y)\delta_{a\bar b}-Y_{a\bar b}$.   Thus, it is sufficient to
consider invariants constructed out of positive powers of $Y$.
One can define a number of different inverses for $Z$ depending on how one
sums over the repeated indices.  Nevertheless, a similar
conclusion holds and one can restrict considerations to invariants
involving positive powers of $Z$.

As a final application of these techniques, we
briefly discuss the $\mathbold{Z_2}$ discrete symmetry $\Phi_1\to
+\Phi_1$, $\Phi_2\to -\Phi_2$,  which implies that
$m_{12}^2=\lam_6=\lam_7=0$ in some basis.  In \sect{sec:discreteB},
we have exhibited U(2)-invariant (basis-independent) conditions
in terms of commutators of Higgs potential coupling second-rank tensors
that imply the existence of a basis where the $\mathbold{Z_2}$
discrete symmetry is manifest.  Here, we provide another method for
constructing U(2)-invariant conditions.  In this method, we introduce
the two orthonormal eigenvectors of
$Y_{ab}$, which we shall denote by $\widehat y$ and $\widehat z$.
It is convenient to define $\widehat z_a\equiv -\epsilon_{ab}y^*_{\bbar}$,
since $\widehat z_a\widehat y^*_{\abar}=0$.
It then follows that:
\beq \label{discrete1}
Y_{a\bbar} \,\widehat y_\abar^\ast\,\widehat z_b = 0\,.
\eeq
That is, $m_{12}^2=0$ in any basis where
$\widehat y=(e^{i\chi_y},0)$.  The freedom to vary $0\leq\chi_y\leq\pi$
corresponds to the simple fact that if $m_{12}^2=0$ in some basis
$\Phi_a$, then $m_{12}^2=0$ is also zero in the rotated basis
$(U_D)_{a\bbar}\Phi_b$, where $U_D={\rm diag}(e^{i\chi_y},e^{-i\chi_y})$.
\Eq{discrete1} also implies that $\widehat y$ and $\widehat z$
transform covariantly with respect to the flavor-SU(2).  To derive
this result, we write $U=e^{i\psi}\widehat U$ as before, where ${\rm
det}~\widehat U=1$.  Using the U(2) transformation law satisfied by
$Y_{a\bbar}$, it follows that $\widehat y$ and $\widehat z$ transform as
$\widehat y\to \widehat U\widehat y$ and
$\widehat z \to \widehat U\widehat z$.
The behavior of $\widehat y_a$ and $\widehat z_a$
under U(1)$_{\rm Y}$ transformations is arbitrary.\footnote{Since
$Y_{a\bbar}$ is invariant with respect to
U(1)$_{\rm Y}$, the eigenvalue equations that defines $\widehat y$ and
$\widehat z$ are unchanged.  The orthonormal eigenvectors are always
defined only up to an arbitrary phase.}  Independently of the behavior of
$\widehat y$ and $\widehat z$ under  U(1)$_{\rm Y}$,
\eq{discrete1} is a U(2)-invariant equation since any overall phase of
$\widehat y_\abar^\ast\,\widehat z_b$ simply drops out of the equation.

\Eq{discrete1} can be used to derive the value of $\cos^2 2\beta$,
where $\widehat v=(\cb\,,\,\sb e^{i\xi})$ in the
eigenbasis of $Y$ (\textit{i.e.}, the basis where $Y$ is diagonal).
In this basis, we may take $\widehat y=(1,0)$ and $\widehat z=(0,1)$.
Transforming to the Higgs basis [\eq{higgsbasis}], one finds that
$\widehat y=(e^{i\chi}\cb\,,\,-e^{i(\xi-\chi)}\sb)$\,,\,\,\,$\widehat z
=(e^{-i(\xi-\chi)} \sb\,,\, e^{-i\chi} \cb)$ and $\widehat
v=(e^{i\chi}\,,\,0)$.  Thus, evaluating \eq{discrete1} in the Higgs basis,
we obtain:
\beq \label{discrete1a}
(Y_{22}-Y_{11})\sb\cb=Y_{12}e^{i(\xi-2\chi)}\cb^2-
Y^*_{12}e^{-i(\xi-2\chi)}\sb^2\,.
\eeq
However, \eq{mab} implies that $Y_{12}e^{i(\xi-2\chi)}$ is real.
Thus, we may write $Y_{12}e^{i(\xi-2\chi)}=\pm |Y_{12}|$, where
the choice of sign reflects two possible transformations from the
eigenbasis of $Y$ to the Higgs basis (which differ by the interchange of
$\Phi_1$ and $\Phi_2$ in the eigenbasis of $Y$).  Consequently,
\eq{discrete1a} implies that
\beq
\frac{c_{2\beta}^2}{s_{2\beta}^2}=\frac{(Y_{11}-Y_{22})^2}{4|Y_{12}|^2}\,,
\eeq
which is equivalent to the result previously given
for $\cos^2 2\beta$ [see \eq{invbeta}] evaluated in the Higgs basis.

The basis-independent conditions for the $\Phi_1\to
+\Phi_1$, $\Phi_2\to -\Phi_2$ discrete symmetry are given by:
\beqa
&&Z_{a\bbar c\dbar}\,\widehat y_\abar^\ast\, \widehat y_b\,
\widehat y_\cbar^\ast \,\widehat z_d=0
\,,\label{discrete2}\\
&&Z_{a\bbar c\dbar}\,\widehat z_\abar^\ast\, \widehat z_b
\,\widehat z_\cbar^\ast\, \widehat y_d=0
\,.\label{discrete3}
\eeqa
As above, any overall phase
redefinitions of $\widehat y$ and $\widehat z$
have no effect on \eqs{discrete2}{discrete3}, which
ensures that these are U(2)-invariant conditions.

One can use these results to obtain conditions that depend only
on invariant combinations of Higgs potential parameters.  As above,
our strategy is to evaluate \eqs{discrete2}{discrete3} in the Higgs
basis.  The outcome of this computation
is two independent relations among invariant combinations of Higgs basis
parameters.  The details are not very illuminating, so we omit them
here.  One can check that evaluating the commutator conditions given
in \sect{sec:discreteB} in the Higgs basis produces equivalent results.

Suppose that in the basis where the discrete symmetry
$\Phi_1\to +\Phi_1$, $\Phi_2\to -\Phi_2$ is manifest, one also has
$\lam_5=0$.  Then, the discrete symmetry is promoted to a global U(1)
Peccei-Quinn symmetry corresponding to $\Phi_1\to e^{i\alpha}\Phi_1$,
$\Phi_2\to e^{-i\alpha}\Phi_2$.  This symmetry is spontaneously broken
by the scalar field vacuum expectation values.  Consequently, the
(tree-level) mass of the CP-odd Higgs scalar (which is to be
identified with the axion~\cite{axion}) must vanish.
The basis-independent conditions
for the existence of the Peccei-Quinn symmetry are given by
\eqst{discrete1}{discrete3} supplemented by a fourth condition:
\beq \label{discrete4}
Z_{a\bbar c\dbar}\,\widehat y_\abar^\ast \,\widehat z_b \,
\widehat y_\cbar^\ast\, \widehat z_d=0
\,.
\eeq
Again, we may evaluate \eq{discrete4} in the Higgs basis.
Combining this result with the corresponding relations obtained
from \eqst{discrete1}{discrete3}, it is possible to prove [using
\eqs{chhiggsmass}{massmha}] that
$\mha^2=0$, which identifies $\ha$ as the axion.

\section{Discovering a Discrete Symmetry}
\label{app:B}

In \Ref{rebelo}, a 2HDM is presented that possesses
a permutation symmetry in which the scalar
Lagrangian is invariant under the interchange of
$\Phi_1$ and $\Phi_2$ in the generic basis.
As a result of the permutation
symmetry, it follows that\footnote{The
authors of \Ref{rebelo} assume in addition that
$\lambda_6$ is real, although this is not required by the permutation
symmetry.}
\beq \label{parmconds}
m_{11}^2=m_{22}^2\,, \lambda_1=\lambda_2\,,~{\rm and}~\lambda_7
=\lambda_6^*\,,~{\rm with}~m_{12}^2~{\rm and}~\lambda_5~{\rm real}\,.
\eeq

In this model, we shall verify that the commutator conditions
[\eqs{comm1}{comm2}] of \sect{sec:discreteB} are satisfied, thereby
``discovering'' that this model respects the discrete symmetry
$\Phi'_1\to\Phi'_1$ and $\Phi'_2\to -\Phi'_2$ in some new basis $\Phi'$.
Inserting \eq{parmconds} into the definitions of $Y$, $Z^{(1)}$ and
$Z^{(11)}$, one can check that the two diagonal elements of each of
the three
matrices are equal.  Similarly, the two off-diagonal elements of each
of the three matrices are equal.  All $2\times 2$ matrices with
these properties commute.  Thus, \eqs{comm1}{comm2} are verified.

Although the invariant conditions provide a powerful method for analysis,
this particular model is simple enough for pedestrian methods.
In particular, starting from
from the $\Phi$-basis that respects the $\Phi_1\tto\Phi_2$
permutation symmetry, one may employ
\eq{utransform} to transform to a new $\Phi'$-basis.   Then, the
new basis respects
the $\Phi'_1\to \Phi'_1$, $\Phi'_2\to -\Phi'_2$ discrete
symmetry if $\theta=\pi/4$ and $\zeta+\gamma=0$.  One can now
check that for $\xi=0$, the $\Phi'$-basis corresponds to the Higgs
basis, while for $\xi\neq 0$ and $\lambda_6$ real,
the $\Phi'$-basis is related to the Higgs basis by a rotation of
$\beta=\xi/2$ or $\beta=(\pi-\xi)/2$.

Having identified the $\mathbold{Z_2}$ discrete symmetry, it follows
that the model is CP-conserving.  This conclusion is less transparent
in the original basis [where \eq{parmconds} is satisfied].
Consider that one
solution to the scalar potential minimum conditions
[\eqs{minconditionsa}{minconditionsb}] is $\beta=\pi/4$.
By subtracting the two minimum conditions, one finds that
\beq \label{minperm}
\left[m_{12}^2-\half v^2(\lambda_5\cos\xi+\lambda_6)\right]\sin\xi=0\,.
\eeq
Two cases can be examined.
If $\Im~\lambda_6\neq 0$, then the only solution to \eq{minperm} is
$\xi=0~({\rm mod}~\pi)$.  If $\lambda_6$ is real, then a second
solution for $\xi$ exists, $\cos\xi=(2m_{12}^2-\lambda_6
v^2)/(\lambda_5 v^2)$, which implies that the vacuum expectation value
has a non-trivial complex phase.  In this case,
one might naively assume that CP is spontaneously broken, but
\Ref{rebelo} shows that this is a false conclusion.

It is noteworthy that in first case above,
$\lambda_6$ was complex in a basis where all other scalar potential
parameters and the vacuum expectation value are real, and in
the second case above,
the vacuum expectation values exhibited a relative phase in a basis
where all Higgs potential parameters are real.
Nevertheless, it is straightforward to check that
in both cases, the three CP-odd invariants
$I_1$, $I_2$ and $I_3$ [\eq{cpoddinv}] are real.  Indeed a
2HDM model that respects a permutation symmetry $\Phi_1\tto\Phi_2$ is
explicitly CP-conserving, and CP is not spontaneously broken by the vacuum.


\end{document}